\documentstyle[aps,epsf,eqsecnum,feynmp
]{revtex}
\newcommand{\setval}{\fmfset{wiggly_len}{1.5mm}\fmfset{arrow_len}{1.5mm}
\fmfset{arrow_ang}{13}\fmfset{dash_len}{1.5mm}\fmfpen{0.125mm}
\fmfset{dot_size}{1thick}}
\newcommand{\dphi}[3]{\frac{\delta #1}{\delta
\parbox{10mm}{\centerline{
\begin{fmfgraph*}(5,3)
\setval
\fmfleft{v1}
\fmfright{v2}
\fmf{plain}{v2,v1}
\fmfv{decor.size=0,label=${\scriptstyle #2}$,l.dist=1mm}{v1}
\fmfv{decor.size=0,label=${\scriptstyle #3}$,l.dist=1mm}{v2}
\end{fmfgraph*}
}}}}
\newcommand{\ddphi}[1]{\frac{\delta^2 #1}{\delta
\parbox{10mm}{\centerline{
\begin{fmfgraph*}(5,3)
\setval
\fmfleft{v1}
\fmfright{v2}
\fmf{plain}{v2,v1}
\fmfv{decor.size=0,label=${\scriptstyle 1}$,l.dist=1mm}{v1}
\fmfv{decor.size=0,label=${\scriptstyle 2}$,l.dist=1mm}{v2}
\end{fmfgraph*}
}}\,\delta
\parbox{10mm}{\centerline{
\begin{fmfgraph*}(5,3)
\setval
\fmfleft{v1}
\fmfright{v2}
\fmf{plain}{v2,v1}
\fmfv{decor.size=0,label=${\scriptstyle 3}$,l.dist=1mm}{v1}
\fmfv{decor.size=0,label=${\scriptstyle 4}$,l.dist=1mm}{v2}
\end{fmfgraph*}
}}}}
\newcommand{\dphiwiggly}[3]{\frac{\delta #1}{\delta
\parbox{10mm}{\centerline{
\begin{fmfgraph*}(5,4)
\setval
\fmfleft{v1}
\fmfright{v2}
\fmf{wiggly,width=0.2mm}{v2,v1}
\fmfv{decor.size=0,label=${\scriptstyle #2}$,l.dist=0.5mm}{v1}
\fmfv{decor.size=0,label=${\scriptstyle #3}$,l.dist=0.5mm}{v2}
\end{fmfgraph*}
}}}}
\newcommand{\dphidouble}[3]{\frac{\delta #1}{\delta
\parbox{10mm}{\centerline{
\begin{fmfgraph*}(5,4)
\setval
\fmfleft{v1}
\fmfright{v2}
\fmf{double,width=0.2mm}{v2,v1}
\fmfv{decor.size=0,label=${\scriptstyle #2}$,l.dist=0.5mm}{v1}
\fmfv{decor.size=0,label=${\scriptstyle #3}$,l.dist=0.5mm}{v2}
\end{fmfgraph*}
}}}}
\begin{document}
\begin{fmffile}{glaum1}
\title{Recursive Graphical Construction of Tadpole-Free Feynman 
Diagrams \\ and Their Weights in $\phi^4$-Theory}

\author{A.~Pelster and K.~Glaum}

\address{Institut f\"ur Theoretische Physik, Freie Universit\"at Berlin, \\
Arnimallee 14, D-14195 Berlin, Germany\\
E-mails: pelster@physik.fu-berlin.de, glaum@physik.fu-berlin.de}

\maketitle

\index{tadpole}
\index{Feynman diagrams}
\index{weights}
\index{$\phi^4$-theory}

\maketitle

\begin{abstract}
We review different approaches to the graphical generation of the 
tadpole-free\index{tadpole}
Feynman diagrams\index{Feynman diagrams} 
of the self-energy\index{self-energy} 
and the one-particle irreducible 
four-point function\index{one-particle irreducible four-point function}.
These are needed for calculating the critical 
exponents\index{critical exponents} of the 
euclidean multicomponent scalar $\phi^4$-theory\index{$\phi^4$-theory} 
with renormalization\index{renormalization}
techniques in $d=4-\epsilon$ dimensions.
\end{abstract}

\section{Introduction}
In 1982 Hagen Kleinert\index{KLEINERT, H.} proposed a program for 
systematically constructing all Feynman diagrams\index{Feynman diagrams} 
of a field theory together
with their proper weights by graphically solving a set of functional
differential equations\cite{PELSKleinert1}. 
It relies on considering a Feynman diagram\index{Feynman diagrams}
as a functional of its graphical elements, i.e., its lines and vertices.
Functional derivatives with respect to these 
graphical elements are represented by
removing lines or vertices of a Feynman diagram\index{Feynman diagrams}
in all possible ways. With these graphical operations, the program proceeds
in four steps. First, a nonlinear 
functional differential equation\index{functional differential equation}
for the free energy\index{free energy}
is derived as a consequence of the field equations.
Subsequently, this functional differential 
equation\index{functional differential equation} 
is converted to a 
recursion relation for the loop expansion coefficients of the 
free energy\index{free energy}.
From its graphical solution, the connected vacuum diagrams are constructed.
Finally, all diagrams of $n$-point functions are obtained
by removing lines or vertices from the connected vacuum diagrams.
This program was recently used
to systematically generate all 
Feynman diagrams\index{Feynman diagrams} of QED\cite{PELSQED} and of 
$\phi^4$-theory\index{$\phi^4$-theory} 
in the disordered, symmetric phase\cite{PELSPHI4} and the ordered,
broken-symmetry phase\cite{PELSBoris1,PELSPHI4B}.

The present article reviews a modification of 
this program\cite{PELSKonstantin}. The aim is to construct directly 
the Feynman diagrams\index{Feynman diagrams}
of $n$-point functions which are relevant for the 
renormalization\index{renormalization} of
a field theory. To this end we consider the 
self-energy\index{self-energy} and the 
one-particle irreducible four-point 
function\index{one-particle irreducible four-point function}
of the euclidean multicomponent
scalar $\phi^4$-theory\index{$\phi^4$-theory} 
as functionals of the free correlation function.
As such they obey a closed set of functional 
differential equations\index{functional differential equation} which 
can be turned into 
graphical recursion relations\index{graphical recursion relation}. 
These are solved order
by order in the number of loops, producing all one-particle irreducible
diagrams\index{one-particle irreducible diagrams} 
with their proper weights\index{weights}. A subsequent absorption of all
tadpole\index{tadpole} corrections in the lines leads to modified graphical
recursion relations for the tadpole\index{tadpole}-free 
one-particle irreducible
diagrams\index{one-particle irreducible diagrams} which are
needed for calculating the critical exponents\index{critical exponents} 
with 
renormalization\index{renormalization}
techniques in $d=4-\epsilon$ dimensions. Finally, we elucidate
how our procedure is related to the method of higher 
functional Legendre transformations\index{Legendre transformation} 
which was also investigated
in Ref.\cite{PELSKleinert1}.
\section{Scalar $\phi^4$-Theory\index{$\phi^4$-theory}}
Consider a self-interacting scalar field $\phi$ with $N$ components 
in $d$ euclidean dimensions whose
thermal fluctuations are controlled by the energy functional
\begin{eqnarray}
\label{PELSEF}
E [ \phi ] = \frac{1}{2} \int_{12} G^{-1}_{12} \phi_1 \phi_2 
+ \frac{1}{4!} \int_{1234} V_{1234}  \phi_1 \phi_2 \phi_3 \phi_4 \, .
\end{eqnarray}
In this short-hand notation, the spatial
arguments and tensor indices of the
field $\phi$, the bilocal kernel $G^{-1}$, and the quartic 
interaction $V$ are indicated
by simple number indices, i.e.,
\begin{eqnarray}
1 \equiv \{ x_1 , \alpha_1 \} \, , \,\,\,\, 
\int_1 \equiv \sum_{\alpha_1} \int d^d x_1 \, , \,\,
\phi_1 \equiv \phi_{\alpha_1} ( x_1 ) \, ,  \,\,\,\, 
%\nonumber \\
G^{-1}_{12} \equiv G^{-1}_{\alpha_1 , \alpha_2} ( x_1 , x_2 ) \, , \,\,\,\,\, 
V_{1234} \equiv V_{\alpha_1 , \alpha_2 , \alpha_3 , \alpha_4} 
( x_1 , x_2 , x_3 , x_4 ) \, .
\end{eqnarray}
The kernel is a functional matrix $G^{-1}$, while the interaction
$V$ is a functional tensor, both being symmetric in their respective
indices. The energy functional (\ref{PELSEF}) describes 
$d$-dimensional euclidean $\phi^4$-theories generically. These are models for
a family of universality classes of continuous 
phase transitions\index{phase transition},
such as the $O(N)$-symmetric $\phi^4$-theory\index{$\phi^4$-theory} 
which serves to derive
the critical phenomena\index{critical phenomena} 
in dilute polymer solutions ($N=0$), Ising- and
Heisenberg-like magnets ($N=1,3$), and superfluids ($N=2$).
In all these
cases, the energy functional (\ref{PELSEF}) is specified by the bilocal kernel
\begin{eqnarray}
G_{\alpha_1 , \alpha_2}^{-1} ( x_1 , x_2 ) & = &
\delta_{\alpha_1 , \alpha_2} \, \left( - \partial_{x_1}^2 + m^2 
\right) \delta ( x_1 - x_2 ) \, , \label{PELSPH1} 
\end{eqnarray}
and by the quartic interaction
\begin{eqnarray}
V_{\alpha_1,\alpha_2,\alpha_3,\alpha_4} ( x_1 , x_2 , x_3 , x_4 ) 
&=&  \frac{g}{3} \, \left\{ 
\delta_{\alpha_1 , \alpha_2} \delta_{\alpha_3 , \alpha_4} +
\delta_{\alpha_1 , \alpha_3} \delta_{\alpha_2 , \alpha_4} +
\delta_{\alpha_1 , \alpha_4} \delta_{\alpha_2 , \alpha_3} \right\} 
%\nonumber \\& & \times 
\delta ( x_1 - x_2 ) \delta ( x_1 - x_3 ) \delta ( x_1 - x_4 ) \, ,
\label{PELSPH2}
\end{eqnarray}
where the bare mass $m^2$ is proportional to the temperature distance
from the critical point, and $g$ denotes the bare coupling constant.
In this article we leave the kernel $G^{-1}$ in the energy functional
(\ref{PELSEF}) completely general, 
except for the symmetry with respect to its indices, and insert
the physical value (\ref{PELSPH1}) only at the end. By doing so, we consider all
statistical quantities derived from (\ref{PELSEF})
as functionals of the free correlation function $G$ which is 
the functional inverse of the kernel $G^{-1}$:
\begin{eqnarray}
\label{PELSFP}
\int_{2} G_{12} \, G^{-1}_{23} = \delta_{13} \, .
\end{eqnarray}
This allows us to
introduce functional derivatives with respect to  $G$ whose
basic rule reflects the symmetry of its indices:
\begin{eqnarray}
\label{PELSDR1}
\frac{\delta G_{12}}{\delta G_{34}} = \frac{1}{2} \left\{ 
\delta_{13} \delta_{42} + \delta_{14} \delta_{32} \right\} \, .
\end{eqnarray}
Such functional derivatives are represented graphically by removing one line
from a Feynman diagram\index{Feynman diagrams} in all possible 
ways\cite{PELSKleinert1,PELSQED,PELSPHI4,PELSBoris1,PELSPHI4B}. 
Thereby each line in a Feynman diagram\index{Feynman diagrams} 
represents a free correlation function 
\begin{eqnarray}
\label{PELSLINE}
\parbox{20mm}{\centerline{
\begin{fmfgraph*}(8,3)
\setval
\fmfleft{v1}
\fmfright{v2}
\fmf{plain}{v1,v2}
\fmflabel{${\scriptstyle 1}$}{v1}
\fmflabel{${\scriptstyle 2}$}{v2}
\end{fmfgraph*}
}} 
\equiv \hspace*{2mm} G_{12} \, ,
\end{eqnarray}
and each vertex represents an integral over the interaction 
\begin{eqnarray}
\label{PELSVERTEX}
\parbox{17mm}{\begin{center}
\begin{fmfgraph*}(6,5)
\setval
\fmfstraight
\fmfleft{i2,i1}
\fmfright{o2,o1}
\fmf{plain}{i1,v1}
\fmf{plain}{v1,i2}
\fmf{plain}{v1,o1}
\fmf{plain}{o2,v1}
\fmfdot{v1}
\end{fmfgraph*}
\end{center}}
\equiv \hspace*{0.2cm} - \, \int_{1234} \, V_{1234} \, .
\end{eqnarray}
Thus the differentiation (\ref{PELSDR1}) is illustrated graphically as
\begin{eqnarray}
\dphi{}{3}{4} 
\parbox{20mm}{\centerline{
\begin{fmfgraph*}(8,3)
\setval
\fmfleft{v1}
\fmfright{v2}
\fmf{plain}{v1,v2}
\fmflabel{${\scriptstyle 1}$}{v1}
\fmflabel{${\scriptstyle 2}$}{v2}
\end{fmfgraph*}
}} = \hspace*{0.2cm} \frac{1}{2} \hspace*{0.2cm} \Bigg\{\hspace*{0.2cm}
\parbox{6mm}{\centerline{
\begin{fmfgraph*}(4,3)
\setval
\fmfforce{0w,0.5h}{i1}
\fmfforce{1w,0.5h}{o1}
\fmfforce{0.5w,0.5h}{v1}
\fmf{plain}{i1,v1}
\fmf{plain}{v1,o1}
\fmfv{decor.size=0, label=${\scriptstyle 1}$, l.dist=1mm, l.angle=-180}{i1}
\fmfv{decor.size=0, label=${\scriptstyle 3}$, l.dist=1mm, l.angle=0}{o1}
\fmfv{decor.shape=circle,decor.filled=empty,decor.size=0.6mm}{v1}
\end{fmfgraph*}}}  \quad
\parbox{6mm}{\centerline{
\begin{fmfgraph*}(4,3)
\setval
\fmfforce{0w,0.5h}{i1}
\fmfforce{1w,0.5h}{o1}
\fmfforce{0.5w,0.5h}{v1}
\fmf{plain}{i1,v1}
\fmf{plain}{v1,o1}
\fmfv{decor.size=0, label=${\scriptstyle 4}$, l.dist=1mm, l.angle=-180}{i1}
\fmfv{decor.size=0, label=${\scriptstyle 2}$, l.dist=1mm, l.angle=0}{o1}
\fmfv{decor.shape=circle,decor.filled=empty,decor.size=0.6mm}{v1}
\end{fmfgraph*}}} \quad + \quad
\parbox{6mm}{\centerline{
\begin{fmfgraph*}(4,3)
\setval
\fmfforce{0w,0.5h}{i1}
\fmfforce{1w,0.5h}{o1}
\fmfforce{0.5w,0.5h}{v1}
\fmf{plain}{i1,v1}
\fmf{plain}{v1,o1}
\fmfv{decor.size=0, label=${\scriptstyle 1}$, l.dist=1mm, l.angle=-180}{i1}
\fmfv{decor.size=0, label=${\scriptstyle 4}$, l.dist=1mm, l.angle=0}{o1}
\fmfv{decor.shape=circle,decor.filled=empty,decor.size=0.6mm}{v1}
\end{fmfgraph*}}} \quad
\parbox{6mm}{\centerline{
\begin{fmfgraph*}(4,3)
\setval
\fmfforce{0w,0.5h}{i1}
\fmfforce{1w,0.5h}{o1}
\fmfforce{0.5w,0.5h}{v1}
\fmf{plain}{i1,v1}
\fmf{plain}{v1,o1}
\fmfv{decor.size=0, label=${\scriptstyle 3}$, l.dist=1mm, l.angle=-180}{i1}
\fmfv{decor.size=0, label=${\scriptstyle 2}$, l.dist=1mm, l.angle=0}{o1}
\fmfv{decor.shape=circle,decor.filled=empty,decor.size=0.6mm}{v1}
\end{fmfgraph*}}} 
\hspace*{0.3cm} \Bigg\} \, ,
\end{eqnarray}
where the elements of Feynman diagrams\index{Feynman diagrams} are extended
by an open dot with two labeled
line ends representing the delta function:
\begin{eqnarray}
\parbox{6mm}{\centerline{
\begin{fmfgraph*}(4,3)
\setval
\fmfforce{0w,0.5h}{i1}
\fmfforce{1w,0.5h}{o1}
\fmfforce{0.5w,0.5h}{v1}
\fmf{plain}{i1,v1}
\fmf{plain}{v1,o1}
\fmfv{decor.size=0, label=${\scriptstyle 1}$, l.dist=1mm, l.angle=-180}{i1}
\fmfv{decor.size=0, label=${\scriptstyle 2}$, l.dist=1mm, l.angle=0}{o1}
\fmfv{decor.shape=circle,decor.filled=empty,decor.size=0.6mm}{v1}
\end{fmfgraph*}}} 
\quad  = \quad \delta_{12} \, .
\end{eqnarray}
\section{Connected Vacuum Diagrams}
By using natural units in which the Boltzmann constant $k_B$ times the
temperature $T$ equals unity, the partition function 
is determined as a functional integral over the Boltzmann
weight $e^{- E [ {\bf \phi} ]}$, i.e.
\begin{eqnarray}
\label{PELSPF}
Z = \int {\cal D} {\bf \phi} \, e^{- E [ {\bf \phi} ]} \, ,
\end{eqnarray}
and may be evaluated perturbatively
as a power series in the interaction $V$.
>From this we obtain
the negative free energy\index{free energy} $W = \ln Z$ as an expansion
\begin{eqnarray}
\label{PELSEX}
W = \sum_{l = 1}^{\infty} W^{(l)} \, ,
\end{eqnarray}
where the coefficients $W^{(l)}$ for each loop order $l \ge 2$
may be displayed for as
connected vacuum diagrams constructed from the lines (\ref{PELSLINE})
and the vertices (\ref{PELSVERTEX}). As has been elaborated in detail in 
Ref.\cite{PELSPHI4}, the connected vaccum diagrams contributing to $W^{(l)}$
follow together with their weights\index{weights} from a 
graphical recursion relation\index{graphical recursion relation} 
which can be written diagrammatically for $l \ge 2$ as
\begin{eqnarray}
W^{(l+1)} =\, \frac{1}{6\,l} \,\,\, \ddphi{W^{(l)}}\,
\parbox{14mm}{\begin{center}
\begin{fmfgraph*}(3,10)
\setval
\fmfstraight
\fmfleft{i1,i2,i3,i4}
\fmfright{o1}
\fmf{plain}{o1,i1}
\fmf{plain}{o1,i2}
\fmf{plain}{i3,o1}
\fmf{plain}{i4,o1}
\fmfdot{o1}
\fmfv{decor.size=0, label=${\scriptstyle 4}$, l.dist=1mm, l.angle=-180}{i1}
\fmfv{decor.size=0, label=${\scriptstyle 3}$, l.dist=1mm, l.angle=-180}{i2}
\fmfv{decor.size=0, label=${\scriptstyle 2}$, l.dist=1mm, l.angle=-180}{i3}
\fmfv{decor.size=0, label=${\scriptstyle 1}$, l.dist=1mm, l.angle=-180}{i4}
\end{fmfgraph*}
\end{center}}
+ \,\, \frac{1}{2\,l} \,\,\, \dphi{W^{(l)}}{1}{2} \,
\parbox{17mm}{\begin{center}
\begin{fmfgraph*}(8,5)
\setval
\fmfstraight
\fmfforce{0w,0h}{i1}
\fmfforce{0w,1h}{i2}
\fmfforce{3/8w,0.5h}{v1}
\fmfforce{1w,0.5h}{v2}
\fmf{plain}{i1,v1}
\fmf{plain}{v1,i2}
\fmf{plain,left}{v1,v2,v1}
\fmfdot{v1}
\fmfv{decor.size=0, label=${\scriptstyle 2}$, l.dist=1mm, l.angle=-180}{i1}
\fmfv{decor.size=0, label=${\scriptstyle 1}$, l.dist=1mm, l.angle=-180}{i2}
\end{fmfgraph*}
\end{center}}
%
%\nonumber \\& & 
\, + \,\, \frac{1}{6\,l} \, \sum\limits_{k=1}^{l-2} 
\, \dphi{W^{(l-k)}}{1}{2}
\parbox{14mm}{\begin{center}
\begin{fmfgraph*}(6,5)
\setval
\fmfstraight
\fmfleft{i2,i1}
\fmfright{o2,o1}
\fmf{plain}{i1,v1}
\fmf{plain}{v1,i2}
\fmf{plain}{v1,o1}
\fmf{plain}{o2,v1}
\fmfdot{v1}
\fmfv{decor.size=0, label=${\scriptstyle 1}$, l.dist=1mm, l.angle=-180}{i1}
\fmfv{decor.size=0, label=${\scriptstyle 2}$, l.dist=1mm, l.angle=-180}{i2}
\fmfv{decor.size=0, label=${\scriptstyle 3}$, l.dist=1mm, l.angle=0}{o1}
\fmfv{decor.size=0, label=${\scriptstyle 4}$, l.dist=1mm, l.angle=0}{o2}
\end{fmfgraph*}
\end{center}}
\dphi{W^{(k+1)}}{3}{4} \, .
\label{PELSGRR}
\end{eqnarray}
This is iterated starting from the two-loop contribution
\begin{eqnarray}
\label{PELSSTTT}
W^{(2)} \, = \, \frac{1}{8} \,\,\,
\parbox{12mm}{\begin{center}
\begin{fmfgraph*}(10,5)
\setval
\fmfleft{i1}
\fmfright{o1}
\fmf{plain,left=1}{i1,v1,i1}
\fmf{plain,left=1}{o1,v1,o1}
\fmfdot{v1}
\end{fmfgraph*}\end{center}} \, .
\end{eqnarray}
The right-hand side of (\ref{PELSGRR}) contains
three different graphical
operations. The first two are linear and involve
one or two line amputations of the previous perturbative order. 
The third operation is nonlinear and mixes two 
different one-line amputations of lower 
orders.\footnote{Note that the first two operations in (\ref{PELSGRR})
were already used in Ref.\cite{PELSNeu} as heuristic algorithms to
generate all topological different connected vacuum diagrams
of the $\phi^4$-theory up to $l=8$ loops with a computer program.
Their corresponding weights were then determined by combinatorial
means with a second computer program.}

The connected vacuum diagrams resulting from the graphical recursion
relation\index{graphical recursion relation} 
(\ref{PELSGRR}) together with their weights\index{weights}
are shown up to $l=5$ loops in Ref.\cite{PELSPHI4}. 
There we observed that the nonlinear operation in (\ref{PELSGRR}) 
does not lead to topologically new diagrams. It only corrects the 
weights\index{weights} generated from the first two linear
operations. Continuing
the solution of the 
graphical recursion relation\index{graphical recursion relation} 
(\ref{PELSGRR}) to higher
loops is an arduous task. We have therefore automatized
the procedure in Ref.\cite{PELSPHI4} by
computer algebra with the help of a unique matrix notation for Feynman
diagrams\index{Feynman diagrams}. 
The corresponding MATHEMATICA program and higher-order
results up to $l=7$ are available on the internet\cite{PELSWWW}.
\section{One-Particle Irreducible 
Diagrams\index{one-particle irreducible diagrams}}
Once the connected vacuum diagrams are known,
the diagrams of the fully-interacting two-point function
\begin{eqnarray}
\label{PELS2P}
\mbox{\boldmath $G$}_{12} = \frac{1}{Z}\,
\int {\cal D} \phi \, \phi_1 \phi_2 \, e^{- E [ \phi ]} 
\end{eqnarray}
and the connected four-point function
\begin{eqnarray}
\label{PELS4P}
\mbox{\boldmath $G$}_{1234}^{\rm c} = \frac{1}{Z}\,
\int {\cal D} \phi \, \phi_1 \phi_2 \phi_3 \phi_4 \, e^{- E [ \phi ]} 
- \mbox{\boldmath $G$}_{12} \mbox{\boldmath $G$}_{34}
- \mbox{\boldmath $G$}_{13} \mbox{\boldmath $G$}_{24}
- \mbox{\boldmath $G$}_{14} \mbox{\boldmath $G$}_{23}
\end{eqnarray}
follow from removing one or two lines in all possible ways,
respectively\cite{PELSPHI4,PELSWWW}:
\begin{eqnarray}
\mbox{\boldmath $G$}_{12} & = & 2 \int_{34} G_{13} G_{24}
\, \frac{\delta W}{\delta G_{34}} \, , \\
\mbox{\boldmath $G$}_{1234}^{\rm c} & = & 2 \int_{56} G_{35} G_{46}
\, \frac{\delta \mbox{\boldmath $G$}_{12}}{\delta G_{56}}
- \mbox{\boldmath $G$}_{13} \mbox{\boldmath $G$}_{24}
- \mbox{\boldmath $G$}_{14} \mbox{\boldmath $G$}_{23} \, ,
\label{PELSGC4}
\end{eqnarray}
Note that removing a line from the diagrams of 
$\mbox{\boldmath $G$}_{12}$ according to the first term of 
(\ref{PELSGC4}) leads to disconnected diagrams which are canceled by the
second and the third term to yield the connected diagrams contributing to
$\mbox{\boldmath $G$}_{1234}^{\rm c}$.
Dropping all diagrams of $\mbox{\boldmath $G$}_{12}$ and
$\mbox{\boldmath $G$}_{1234}^{\rm c}$ which would fall into two
pieces by removing a line, one obtains the one-particle
irreducible diagrams\index{one-particle irreducible diagrams} 
of the self-energy\index{self-energy}
\begin{eqnarray}
\label{PELSSE} 
\Sigma_{12} = G_{12}^{-1} - \mbox{\boldmath $G$}_{12}^{-1}\, ,
\end{eqnarray}
with $\mbox{\boldmath $G$}^{-1}$ being the functional inverse
of $\mbox{\boldmath $G$}$
\begin{eqnarray}
\label{PELSFPC}
\int_{2} \mbox{\boldmath $G$}_{12} 
\,  \mbox{\boldmath $G$}_{23}^{-1} = \delta_{13} \, ,
\end{eqnarray}
and the one-particle irreducible four-point 
function\index{one-particle irreducible four-point function}
\begin{eqnarray}
\label{PELSIRF}
\Gamma_{1234} = - \int_{5678} 
\mbox{\boldmath $G$}^{-1}_{15}
\mbox{\boldmath $G$}^{-1}_{26}
\mbox{\boldmath $G$}^{-1}_{37}
\mbox{\boldmath $G$}^{-1}_{48}
\mbox{\boldmath $G$}^{{\rm c}}_{5678} \, .
\end{eqnarray}
Along these lines all diagrams and their weights\index{weights} 
were constructed which
are relevant for the five-loop renormalization\index{renormalization} of the 
$\phi^4$-theory\index{$\phi^4$-theory}
in $d=4-\epsilon$ dimensions\cite{PELSNeu,PELSFIVE,PELSVerena}.

At this stage the question arises whether the 
one-particle irreducible diagrams\index{one-particle irreducible diagrams}
can also be generated in a direct graphical way without any reference
to the connected vacuum diagrams. This is indeed possible and will be
elaborated in detail in Ref.\cite{PELSKonstantin}. Here we restrict ourselves to
present some of the results. To this end we extend the elements of Feynman
diagrams\index{Feynman diagrams} 
by a double straight line representing the fully-interacting two-point 
function
\begin{eqnarray}
\parbox{10mm}{\centerline{
\begin{fmfgraph*}(7,3)
\setval
\fmfforce{0w,1/2h}{v1}
\fmfforce{1w,1/2h}{v2}
\fmf{double,width=0.2mm}{v2,v1}
\fmfv{decor,size=0, label=${\scriptstyle 1}$, l.dist=1mm, l.angle=-180}{v1}
\fmfv{decor,size=0, label=${\scriptstyle 2}$, l.dist=1mm, l.angle=0}{v2}
\end{fmfgraph*} } } 
\hspace*{0.2cm} = \hspace*{0.2cm} \mbox{\boldmath $G$}_{12} \, ,
\end{eqnarray}
and by a $2$- and $4$-vertex with a big open dot representing the 
self-energy\index{self-energy}
and the one-particle irreducible four-point 
function\index{one-particle irreducible four-point function}

\begin{eqnarray}
\parbox{11mm}{\centerline{
\begin{fmfgraph*}(9,3)
\setval
\fmfforce{0w,1/2h}{v1}
\fmfforce{3/9w,1/2h}{v2}
\fmfforce{6/9w,1/2h}{v3}
\fmfforce{1w,1/2h}{v4}
\fmfforce{1/2w,1h}{v5}
\fmfforce{1/2w,0h}{v6}
\fmf{plain,width=0.2mm}{v1,v2}
\fmf{plain,width=0.2mm}{v3,v4}
\fmf{double,width=0.2mm,left=1}{v5,v6,v5}
\fmfv{decor.size=0, label=${\scriptstyle 1}$, l.dist=1mm, l.angle=-180}{v1}
\fmfv{decor.size=0, label=${\scriptstyle 2}$, l.dist=1mm, l.angle=0}{v4}
\end{fmfgraph*} } } 
\hspace*{0.3cm} & = & \hspace*{0.3cm} \Sigma_{12} \, ,  \\ && \nonumber \\
\parbox{10mm}{\centerline{
\begin{fmfgraph*}(6,6)
\setval
\fmfforce{0w,0h}{v1}
\fmfforce{0w,1h}{v2}
\fmfforce{1w,1h}{v3}
\fmfforce{1w,0h}{v4}
\fmfforce{1/3w,1/3h}{v5}
\fmfforce{1/3w,2/3h}{v6}
\fmfforce{2/3w,2/3h}{v7}
\fmfforce{2/3w,1/3h}{v8}
\fmf{plain}{v1,v5}
\fmf{plain}{v2,v6}
\fmf{plain}{v3,v7}
\fmf{plain}{v4,v8}
\fmf{double,width=0.2mm,left=1}{v5,v7,v5}
\fmfv{decor.size=0, label=${\scriptstyle 1}$, l.dist=1mm, l.angle=-135}{v1}
\fmfv{decor.size=0, label=${\scriptstyle 2}$, l.dist=1mm, l.angle=135}{v2}
\fmfv{decor.size=0, label=${\scriptstyle 3}$, l.dist=1mm, l.angle=45}{v3}
\fmfv{decor.size=0, label=${\scriptstyle 4}$, l.dist=1mm, l.angle=-45}{v4}
\end{fmfgraph*} } } 
\hspace*{0.3cm} & = & \hspace*{0.3cm} \Gamma_{1234} \, , \\ && \nonumber
\end{eqnarray}
respectively.
It turns out that the self-energy\index{self-energy} 
$\Sigma$ follows from an integral
equation which reads graphically
\begin{eqnarray}
\label{PELSSELF}
\parbox{11mm}{\centerline{
\begin{fmfgraph*}(9,3)
\setval
\fmfforce{0w,1/2h}{v1}
\fmfforce{3/9w,1/2h}{v2}
\fmfforce{6/9w,1/2h}{v3}
\fmfforce{1w,1/2h}{v4}
\fmfforce{1/2w,1h}{v5}
\fmfforce{1/2w,0h}{v6}
\fmf{plain,width=0.2mm}{v1,v2}
\fmf{plain,width=0.2mm}{v3,v4}
\fmf{double,width=0.2mm,left=1}{v5,v6,v5}
\fmfv{decor.size=0, label=${\scriptstyle 1}$, l.dist=1mm, l.angle=-180}{v1}
\fmfv{decor.size=0, label=${\scriptstyle 2}$, l.dist=1mm, l.angle=0}{v4}
\end{fmfgraph*} } } 
\hspace*{3mm} = \hspace*{2mm} \frac{1}{2} \hspace*{3mm}
\parbox{9mm}{\centerline{
\begin{fmfgraph*}(6,5)
\setval
\fmfforce{0w,0h}{v1}
\fmfforce{1/2w,0h}{v2}
\fmfforce{1w,0h}{v3}
\fmfforce{1/12w,1/2h}{v4}
\fmfforce{11/12w,1/2h}{v5}
\fmf{plain}{v1,v3}
\fmf{double,width=0.2mm,left=1}{v4,v5,v4}
\fmfv{decor.size=0, label=${\scriptstyle 1}$, l.dist=1mm, l.angle=-180}{v1}
\fmfv{decor.size=0, label=${\scriptstyle 2}$, l.dist=1mm, l.angle=0}{v3}
\fmfdot{v2}
\end{fmfgraph*} } }
\hspace*{3mm} + \hspace*{2mm} \frac{1}{6} \hspace*{3mm}
\parbox{17mm}{\centerline{
\begin{fmfgraph*}(14,6)
\setval
\fmfforce{0w,1/2h}{v1}
\fmfforce{3/14w,1/2h}{v2}
\fmfforce{8/14w,1/2h}{v3}
\fmfforce{11/14w,1/2h}{v4}
\fmfforce{1w,1/2h}{v5}
\fmfforce{9.2/14w,2.2/3h}{v6}
\fmfforce{9.2/14w,0.8/3h}{v7}
\fmf{plain}{v1,v2}
\fmf{plain}{v4,v5}
\fmf{double,width=0.2mm,left=0.8}{v2,v6}
\fmf{double,width=0.2mm}{v2,v3}
\fmf{double,width=0.2mm,right=0.8}{v2,v7}
\fmf{double,width=0.2mm,right=1}{v3,v4,v3}
\fmfv{decor.size=0, label=${\scriptstyle 1}$, l.dist=1mm, l.angle=-180}{v1}
\fmfv{decor.size=0, label=${\scriptstyle 2}$, l.dist=1mm, l.angle=0}{v5}
\fmfdot{v2}
\end{fmfgraph*} } } \hspace*{0.3cm} .
\end{eqnarray}
Thereby the connected two-point function 
$\mbox{\boldmath $G$}$ is obtained from (\ref{PELSSE}) 
according to the Dyson equation
\begin{eqnarray}
\label{PELSDY}
\parbox{10mm}{\centerline{
\begin{fmfgraph*}(7,3)
\setval
\fmfforce{0w,1/2h}{v1}
\fmfforce{1w,1/2h}{v2}
\fmf{double,width=0.2mm}{v2,v1}
\fmfv{decor,size=0, label=${\scriptstyle 1}$, l.dist=1mm, l.angle=-180}{v1}
\fmfv{decor,size=0, label=${\scriptstyle 2}$, l.dist=1mm, l.angle=0}{v2}
\end{fmfgraph*} } } 
\hspace*{3mm} = \hspace*{3mm}
\parbox{10mm}{\centerline{
\begin{fmfgraph*}(7,3)
\setval
\fmfforce{0w,1/2h}{v1}
\fmfforce{1w,1/2h}{v2}
\fmf{plain,width=0.2mm}{v2,v1}
\fmfv{decor,size=0, label=${\scriptstyle 1}$, l.dist=1mm, l.angle=-180}{v1}
\fmfv{decor,size=0, label=${\scriptstyle 2}$, l.dist=1mm, l.angle=0}{v2}
\end{fmfgraph*} } } 
\hspace*{3mm} + \hspace*{3mm}
\parbox{15mm}{\centerline{
\begin{fmfgraph*}(13,3)
\setval
\fmfforce{0w,1/2h}{v1}
\fmfforce{5/13w,1/2h}{v2}
\fmfforce{8/13w,1/2h}{v3}
\fmfforce{1w,1/2h}{v4}
\fmfforce{1/2w,1h}{v5}
\fmfforce{1/2w,0h}{v6}
\fmf{plain,width=0.2mm}{v1,v2}
\fmf{double,width=0.2mm}{v3,v4}
\fmf{double,width=0.2mm,left=1}{v5,v6,v5}
\fmfv{decor.size=0, label=${\scriptstyle 1}$, l.dist=1mm, l.angle=-180}{v1}
\fmfv{decor.size=0, label=${\scriptstyle 2}$, l.dist=1mm, l.angle=0}{v4}
\end{fmfgraph*} } } \hspace*{0.3cm} .
\end{eqnarray}
Solving the Eqs.~(\ref{PELSSELF}) and (\ref{PELSDY}) iteratively
necessitates the knowledge of the one-particle irreducible 
four-point function\index{one-particle irreducible four-point function}
$\Gamma$. Its diagrams could be determined from
\begin{eqnarray}
\hspace*{-0.2cm}
\parbox{10mm}{\centerline{
\begin{fmfgraph*}(6,6)
\setval
\fmfforce{0w,0h}{v1}
\fmfforce{0w,1h}{v2}
\fmfforce{1w,1h}{v3}
\fmfforce{1w,0h}{v4}
\fmfforce{1/3w,1/3h}{v5}
\fmfforce{1/3w,2/3h}{v6}
\fmfforce{2/3w,2/3h}{v7}
\fmfforce{2/3w,1/3h}{v8}
\fmf{plain}{v1,v5}
\fmf{plain}{v2,v6}
\fmf{plain}{v3,v7}
\fmf{plain}{v4,v8}
\fmf{double,width=0.2mm,left=1}{v5,v7,v5}
\fmfv{decor.size=0, label=${\scriptstyle 1}$, l.dist=1mm, l.angle=-135}{v1}
\fmfv{decor.size=0, label=${\scriptstyle 2}$, l.dist=1mm, l.angle=135}{v2}
\fmfv{decor.size=0, label=${\scriptstyle 3}$, l.dist=1mm, l.angle=45}{v3}
\fmfv{decor.size=0, label=${\scriptstyle 4}$, l.dist=1mm, l.angle=-45}{v4}
\end{fmfgraph*} } } 
\hspace*{2mm}  &=&  \hspace*{2mm} 2 \hspace*{2mm}
\dphi{
\parbox{14mm}{\centerline{
\begin{fmfgraph*}(9,3)
\setval
\fmfforce{0w,3/4h}{v1}
\fmfforce{3/9w,3/4h}{v2}
\fmfforce{6/9w,3/4h}{v3}
\fmfforce{1w,3/4h}{v4}
\fmfforce{1/2w,5/4h}{v5}
\fmfforce{1/2w,1/4h}{v6}
\fmf{plain}{v1,v2}
\fmf{plain}{v3,v4}
\fmf{double,width=0.2mm,left=1}{v5,v6,v5}
\fmfv{decor.size=0, label=${\scriptstyle 1}$, l.dist=0.5mm, l.angle=-180}{v1}
\fmfv{decor.size=0, label=${\scriptstyle 2}$, l.dist=0.5mm, l.angle=0}{v4}
\end{fmfgraph*} } }
}{3}{4} 
\hspace*{2mm} + \hspace*{2mm} 2 \hspace*{2mm}
\parbox{14mm}{\centerline{
\begin{fmfgraph*}(11,3)
\setval
\fmfforce{0w,1/2h}{v1}
\fmfforce{1w,1/2h}{v2}
\fmfforce{3/11w,1/2h}{v3}
\fmfforce{6/11w,1/2h}{v4}
\fmf{plain}{v1,v3}
\fmf{plain}{v2,v4}  
\fmf{double,width=0.2mm,left=1}{v3,v4,v3}
\fmfv{decor.size=0, label=${\scriptstyle 3}$, l.dist=1mm, l.angle=180}{v1}
\fmfv{decor.size=0, label=${\scriptstyle 5}$, l.dist=1mm, l.angle=0}{v2}
\end{fmfgraph*} } } 
\hspace*{2mm}  
\dphi{
\parbox{14mm}{\centerline{
\begin{fmfgraph*}(9,3)
\setval
\fmfforce{0w,3/4h}{v1}
\fmfforce{3/9w,3/4h}{v2}
\fmfforce{6/9w,3/4h}{v3}
\fmfforce{1w,3/4h}{v4}
\fmfforce{1/2w,5/4h}{v5}
\fmfforce{1/2w,1/4h}{v6}
\fmf{plain}{v1,v2}
\fmf{plain}{v3,v4}
\fmf{double,width=0.2mm,left=1}{v5,v6,v5}
\fmfv{decor.size=0, label=${\scriptstyle 1}$, l.dist=0.5mm, l.angle=-180}{v1}
\fmfv{decor.size=0, label=${\scriptstyle 2}$, l.dist=0.5mm, l.angle=0}{v4}
\end{fmfgraph*} } }
}{5}{6}
\hspace*{2mm} 
\parbox{14mm}{\centerline{
\begin{fmfgraph*}(11,3)
\setval
\fmfforce{0w,1/2h}{v1}
\fmfforce{1w,1/2h}{v2}
\fmfforce{5/11w,1/2h}{v3}
\fmfforce{8/11w,1/2h}{v4}
\fmf{plain}{v1,v3}
\fmf{plain}{v2,v4}  
\fmf{double,width=0.2mm,left=1}{v3,v4,v3}
\fmfv{decor.size=0, label=${\scriptstyle 6}$, l.dist=1mm, l.angle=180}{v1}
\fmfv{decor.size=0, label=${\scriptstyle 4}$, l.dist=1mm, l.angle=0}{v2}
\end{fmfgraph*} } }  
\nonumber \\*[0.2cm]
& & 
\hspace*{2mm} - \hspace*{2mm} 2 \hspace*{2mm} 
\parbox{14mm}{\centerline{
\begin{fmfgraph*}(11,3)
\setval
\fmfforce{0w,1/2h}{v1}
\fmfforce{1w,1/2h}{v2}
\fmfforce{3/11w,1/2h}{v3}
\fmfforce{6/11w,1/2h}{v4}
\fmf{plain}{v1,v3}
\fmf{plain}{v2,v4}  
\fmf{double,width=0.2mm,left=1}{v3,v4,v3}
\fmfv{decor.size=0, label=${\scriptstyle 3}$, l.dist=1mm, l.angle=180}{v1}
\fmfv{decor.size=0, label=${\scriptstyle 5}$, l.dist=1mm, l.angle=0}{v2}
\end{fmfgraph*} } } 
\hspace*{2mm}  
\dphi{
\parbox{14mm}{\centerline{
\begin{fmfgraph*}(9,3)
\setval
\fmfforce{0w,3/4h}{v1}
\fmfforce{3/9w,3/4h}{v2}
\fmfforce{6/9w,3/4h}{v3}
\fmfforce{1w,3/4h}{v4}
\fmfforce{1/2w,5/4h}{v5}
\fmfforce{1/2w,1/4h}{v6}
\fmf{plain}{v1,v2}
\fmf{plain}{v3,v4}
\fmf{double,width=0.2mm,left=1}{v5,v6,v5}
\fmfv{decor.size=0, label=${\scriptstyle 1}$, l.dist=0.5mm, l.angle=-180}{v1}
\fmfv{decor.size=0, label=${\scriptstyle 2}$, l.dist=0.5mm, l.angle=0}{v4}
\end{fmfgraph*} } }
}{5}{4} 
\hspace*{2mm} - \hspace*{2mm} 2 \hspace*{2mm}  
\dphi{
\parbox{14mm}{\centerline{
\begin{fmfgraph*}(9,3)
\setval
\fmfforce{0w,3/4h}{v1}
\fmfforce{3/9w,3/4h}{v2}
\fmfforce{6/9w,3/4h}{v3}
\fmfforce{1w,3/4h}{v4}
\fmfforce{1/2w,5/4h}{v5}
\fmfforce{1/2w,1/4h}{v6}
\fmf{plain}{v1,v2}
\fmf{plain}{v3,v4}
\fmf{double,width=0.2mm,left=1}{v5,v6,v5}
\fmfv{decor.size=0, label=${\scriptstyle 1}$, l.dist=0.5mm, l.angle=-180}{v1}
\fmfv{decor.size=0, label=${\scriptstyle 2}$, l.dist=0.5mm, l.angle=0}{v4}
\end{fmfgraph*} } }
}{3}{6} 
\hspace*{2mm} 
\parbox{14mm}{\centerline{
\begin{fmfgraph*}(11,3)
\setval
\fmfforce{0w,1/2h}{v1}
\fmfforce{1w,1/2h}{v2}
\fmfforce{5/11w,1/2h}{v3}
\fmfforce{8/11w,1/2h}{v4}
\fmf{plain}{v1,v3}
\fmf{plain}{v2,v4}  
\fmf{double,width=0.2mm,left=1}{v3,v4,v3}
\fmfv{decor.size=0, label=${\scriptstyle 6}$, l.dist=1mm, l.angle=180}{v1}
\fmfv{decor.size=0, label=${\scriptstyle 4}$, l.dist=1mm, l.angle=0}{v2}
\end{fmfgraph*} } } \hspace*{0.1cm} \, . 
\label{PELSFOUR}
\end{eqnarray}
However, such a procedure would have one disadvantage: removing a line
in the diagrams of the self-energy\index{self-energy} $\Sigma$ also 
leads to
one-particle reducible diagrams\index{one-particle irreducible diagrams} 
which are later on cancelled by the
third and the fourth term in (\ref{PELSFOUR}). As the number of undesired 
one-particle reducible diagrams\index{one-particle irreducible diagrams} 
occuring at an intermediate step of the
calculation increases with the perturbative order, this procedure is
quite inefficient in determining the diagrams of the one-particle
irreducible four-point 
function\index{one-particle irreducible four-point function}
$\Gamma$. By inserting (\ref{PELSSELF}) 
in (\ref{PELSFOUR}) it turns out that we can derive another
equation for $\Gamma$ whose iterative solution only involves one-particle
irreducible diagrams\index{one-particle irreducible diagrams}:
\begin{eqnarray}
  \parbox{10mm}{\centerline{
  \begin{fmfgraph*}(6,6)
  \setval
  \fmfforce{0w,0h}{v1}
  \fmfforce{0w,1h}{v2}
  \fmfforce{1w,1h}{v3}
  \fmfforce{1w,0h}{v4}
  \fmfforce{1/3w,1/3h}{v5}
  \fmfforce{1/3w,2/3h}{v6}
  \fmfforce{2/3w,2/3h}{v7}
  \fmfforce{2/3w,1/3h}{v8}
  \fmf{plain}{v1,v5}
  \fmf{plain}{v2,v6}
  \fmf{plain}{v3,v7}
  \fmf{plain}{v4,v8}
  \fmf{double,width=0.2mm,left=1}{v5,v7,v5}
  \fmfv{decor.size=0, label=${\scriptstyle 1}$, l.dist=1mm, l.angle=-135}{v1}
  \fmfv{decor.size=0, label=${\scriptstyle 2}$, l.dist=1mm, l.angle=135}{v2}
  \fmfv{decor.size=0, label=${\scriptstyle 3}$, l.dist=1mm, l.angle=45}{v3}
  \fmfv{decor.size=0, label=${\scriptstyle 4}$, l.dist=1mm, l.angle=-45}{v4}
  \end{fmfgraph*} } } 
%%%%
\hspace*{2mm} & = & \hspace*{3mm}
%%%%
  \parbox{7mm}{\centerline{
  \begin{fmfgraph*}(4.5,4.5)
  \setval
  \fmfforce{0w,0h}{v1}
  \fmfforce{0w,1h}{v2}
  \fmfforce{1w,1h}{v3}
  \fmfforce{1w,0h}{v4}
  \fmfforce{1/2w,1/2h}{v5}
  \fmf{plain}{v1,v3}
  \fmf{plain}{v2,v4}
  \fmfv{decor.size=0, label=${\scriptstyle 1}$, l.dist=1mm, l.angle=-135}{v1}
  \fmfv{decor.size=0, label=${\scriptstyle 2}$, l.dist=1mm, l.angle=135}{v2}
  \fmfv{decor.size=0, label=${\scriptstyle 3}$, l.dist=1mm, l.angle=45}{v3}
  \fmfv{decor.size=0, label=${\scriptstyle 4}$, l.dist=1mm, l.angle=-45}{v4}
  \fmfdot{v5}
  \end{fmfgraph*} } } 
%%%%
\hspace*{3mm} + \hspace*{2mm} \frac{1}{3} \hspace*{3mm}
%%%%
  \parbox{10mm}{\centerline{
  \begin{fmfgraph*}(8,8)
  \setval
  \fmfforce{0w,5.5/8h}{v1}
  \fmfforce{3/8w,5.5/8h}{v2}
  \fmfforce{1w,5.5/8h}{v3}
  \fmfforce{1w,1.5/8h}{v4}
  \fmfforce{1w,9.5/8h}{v5}
  \fmf{plain}{v1,v2}
  \fmf{plain}{v2,v3}
  \fmf{plain,right=0.3}{v2,v4}
  \fmf{double,width=0.2mm,left=0.3}{v2,v5}
  \fmfv{decor.size=0, label=${\scriptstyle 1}$, l.dist=1mm, l.angle=-180}{v1}
  \fmfv{decor.size=0, label=${\scriptstyle 6}$, l.dist=1mm, l.angle=0}{v3}
  \fmfv{decor.size=0, label=${\scriptstyle 7}$, l.dist=1mm, l.angle=0}{v4}
  \fmfv{decor.size=0, label=${\scriptstyle 5}$, l.dist=1mm, l.angle=0}{v5}
  \fmfdot{v2}
  \end{fmfgraph*} } } 
%%%%
\hspace*{3mm} 
%%%%
\dphi{
  \parbox{11mm}{\centerline{
  \begin{fmfgraph*}(6,7)
  \setval
  \fmfforce{0w,2/7h}{v1}
  \fmfforce{0w,8/7h}{v2}
  \fmfforce{1w,8/7h}{v3}
  \fmfforce{1w,2/7h}{v4}
  \fmfforce{1/3w,4/7h}{v5}
  \fmfforce{1/3w,6/7h}{v6}
  \fmfforce{2/3w,6/7h}{v7}
  \fmfforce{2/3w,4/7h}{v8}
  \fmf{plain}{v1,v5}
  \fmf{plain}{v2,v6}
  \fmf{plain}{v3,v7}
  \fmf{plain}{v4,v8}
  \fmf{double,width=0.2mm,left=1}{v5,v7,v5}
  \fmfv{decor.size=0, label=${\scriptstyle 5}$, l.dist=0.5mm, l.angle=-180}{v1}
  \fmfv{decor.size=0, label=${\scriptstyle 2}$, l.dist=0.5mm, l.angle=180}{v2}
  \fmfv{decor.size=0, label=${\scriptstyle 3}$, l.dist=0.5mm, l.angle=0}{v3}
  \fmfv{decor.size=0, label=${\scriptstyle 4}$, l.dist=0.5mm, l.angle=0}{v4}
  \end{fmfgraph*} } } 
}{6}{7}
%\nonumber  \\ && \nonumber \\
%%%%
%\hspace*{3mm} 
%&&
+ \hspace*{2mm} \frac{1}{2} \hspace*{3mm}
%%%%
  \parbox{14mm}{\centerline{
  \begin{fmfgraph*}(11,6)
  \setval
  \fmfforce{0w,1/6h}{v1}
  \fmfforce{0w,5/6h}{v2}
  \fmfforce{1w,1h}{v3}
  \fmfforce{1w,0h}{v4}
  \fmfforce{7/11w,1/3h}{v5}
  \fmfforce{7/11w,2/3h}{v6}
  \fmfforce{9/11w,2/3h}{v7}
  \fmfforce{9/11w,1/3h}{v8}
  \fmfforce{2/11w,1/2h}{v9}
  \fmf{plain}{v1,v9}
  \fmf{plain}{v2,v9}
  \fmf{plain}{v3,v7}
  \fmf{plain}{v4,v8}
  \fmf{double,width=0.2mm,left=1}{v5,v7,v5}
  \fmf{double,width=0.2mm,left=0.7}{v9,v6}
  \fmf{double,width=0.2mm,right=0.7}{v9,v5}
  \fmfv{decor.size=0, label=${\scriptstyle 1}$, l.dist=1mm, l.angle=-135}{v1}
  \fmfv{decor.size=0, label=${\scriptstyle 2}$, l.dist=1mm, l.angle=135}{v2}
  \fmfv{decor.size=0, label=${\scriptstyle 3}$, l.dist=1mm, l.angle=25}{v3}
  \fmfv{decor.size=0, label=${\scriptstyle 4}$, l.dist=1mm, l.angle=-25}{v4}
  \fmfdot{v9}
  \end{fmfgraph*} } } 
%%%%
\hspace*{3mm} + \hspace*{2mm} \frac{1}{2} \hspace*{3mm}
%%%%
  \parbox{14mm}{\centerline{
  \begin{fmfgraph*}(11,6)
  \setval
  \fmfforce{0w,1/6h}{v1}
  \fmfforce{0w,5/6h}{v2}
  \fmfforce{1w,1h}{v3}
  \fmfforce{1w,0h}{v4}
  \fmfforce{7/11w,1/3h}{v5}
  \fmfforce{7/11w,2/3h}{v6}
  \fmfforce{9/11w,2/3h}{v7}
  \fmfforce{9/11w,1/3h}{v8}
  \fmfforce{2/11w,1/2h}{v9}
  \fmf{plain}{v1,v9}
  \fmf{plain}{v2,v9}
  \fmf{plain}{v3,v7}
  \fmf{plain}{v4,v8}
  \fmf{double,width=0.2mm,left=1}{v5,v7,v5}
  \fmf{double,width=0.2mm,left=0.7}{v9,v6}
  \fmf{double,width=0.2mm,right=0.7}{v9,v5}
  \fmfv{decor.size=0, label=${\scriptstyle 1}$, l.dist=1mm, l.angle=-135}{v1}
  \fmfv{decor.size=0, label=${\scriptstyle 3}$, l.dist=1mm, l.angle=135}{v2}
  \fmfv{decor.size=0, label=${\scriptstyle 4}$, l.dist=1mm, l.angle=25}{v3}
  \fmfv{decor.size=0, label=${\scriptstyle 2}$, l.dist=1mm, l.angle=-25}{v4}
  \fmfdot{v9}
  \end{fmfgraph*} } } 
%%%%
%%%%                 %%%%%%%%%%%%%%%%%%%%
\hspace*{0.2cm} + \hspace*{2mm} \frac{1}{2} \hspace*{3mm}
%%%%
  \parbox{14mm}{\centerline{
  \begin{fmfgraph*}(11,6)
  \setval
  \fmfforce{0w,1/6h}{v1}
  \fmfforce{0w,5/6h}{v2}
  \fmfforce{1w,1h}{v3}
  \fmfforce{1w,0h}{v4}
  \fmfforce{7/11w,1/3h}{v5}
  \fmfforce{7/11w,2/3h}{v6}
  \fmfforce{9/11w,2/3h}{v7}
  \fmfforce{9/11w,1/3h}{v8}
  \fmfforce{2/11w,1/2h}{v9}
  \fmf{plain}{v1,v9}
  \fmf{plain}{v2,v9}
  \fmf{plain}{v3,v7}
  \fmf{plain}{v4,v8}
  \fmf{double,width=0.2mm,left=1}{v5,v7,v5}
  \fmf{double,width=0.2mm,left=0.7}{v9,v6}
  \fmf{double,width=0.2mm,right=0.7}{v9,v5}
  \fmfv{decor.size=0, label=${\scriptstyle 1}$, l.dist=1mm, l.angle=-135}{v1}
  \fmfv{decor.size=0, label=${\scriptstyle 4}$, l.dist=1mm, l.angle=135}{v2}
  \fmfv{decor.size=0, label=${\scriptstyle 2}$, l.dist=1mm, l.angle=25}{v3}
  \fmfv{decor.size=0, label=${\scriptstyle 3}$, l.dist=1mm, l.angle=-25}{v4}
  \fmfdot{v9}
  \end{fmfgraph*} } } 
%%%%
\nonumber  \\ && \nonumber \\
&& + \hspace*{2mm} \frac{1}{6} \hspace*{2mm}
%%%%
  \parbox{14mm}{\centerline{
  \begin{fmfgraph*}(11,11)
  \setval
  \fmfforce{1/11w,0h}{v1}
  \fmfforce{0w,1h}{v2}
  \fmfforce{3/11w,2/11h}{v3}
  \fmfforce{2/11w,9/11h}{v4}
  \fmfforce{3/11w,6.5/11h}{v5}
  \fmfforce{1.5/11w,7.5/11h}{v6}
  \fmfforce{4.5/11w,8.5/11h}{v7}
  \fmfforce{7/11w,4.5/11h}{v8}
  \fmfforce{7/11w,6.5/11h}{v9}
  \fmfforce{9.2/11w,6.7/11h}{v10}
  \fmfforce{9/11w,4.5/11h}{v11}
  \fmfforce{1w,8.5/11h}{v12}
  \fmfforce{1w,2.5/11h}{v13}
  \fmf{plain}{v1,v3}
  \fmf{plain}{v2,v4}
  \fmf{plain}{v10,v12}
  \fmf{plain}{v11,v13}
  \fmf{double,width=0.2mm,left=1}{v6,v7,v6}
  \fmf{double,width=0.2mm,left=1}{v9,v11,v9}
  \fmf{double,width=0.2mm}{v3,v5}
  \fmf{double,width=0.2mm,left=0.5}{v3,v6}
  \fmf{double,width=0.2mm,left=0.3}{v7,v9}
  \fmf{double,width=0.2mm,right=0.4}{v3,v8}
  \fmfv{decor.size=0, label=${\scriptstyle 1}$, l.dist=1mm, l.angle=-180}{v1}
  \fmfv{decor.size=0, label=${\scriptstyle 2}$, l.dist=1mm, l.angle=180}{v2}
  \fmfv{decor.size=0, label=${\scriptstyle 3}$, l.dist=1mm, l.angle=45}{v12}
  \fmfv{decor.size=0, label=${\scriptstyle 4}$, l.dist=1mm, l.angle=-45}{v13}
  \fmfdot{v3}
  \end{fmfgraph*} } } 
%%%%%
\hspace*{3mm} + \hspace*{2mm} \frac{1}{6} \hspace*{2mm}
%%%%
  \parbox{14mm}{\centerline{
  \begin{fmfgraph*}(11,11)
  \setval
  \fmfforce{1/11w,0h}{v1}
  \fmfforce{0w,1h}{v2}
  \fmfforce{3/11w,2/11h}{v3}
  \fmfforce{2/11w,9/11h}{v4}
  \fmfforce{3/11w,6.5/11h}{v5}
  \fmfforce{1.5/11w,7.5/11h}{v6}
  \fmfforce{4.5/11w,8.5/11h}{v7}
  \fmfforce{7/11w,4.5/11h}{v8}
  \fmfforce{7/11w,6.5/11h}{v9}
  \fmfforce{9.2/11w,6.7/11h}{v10}
  \fmfforce{9/11w,4.5/11h}{v11}
  \fmfforce{1w,8.5/11h}{v12}
  \fmfforce{1w,2.5/11h}{v13}
  \fmf{plain}{v1,v3}
  \fmf{plain}{v2,v4}
  \fmf{plain}{v10,v12}
  \fmf{plain}{v11,v13}
  \fmf{double,width=0.2mm,left=1}{v6,v7,v6}
  \fmf{double,width=0.2mm,left=1}{v9,v11,v9}
  \fmf{double,width=0.2mm}{v3,v5}
  \fmf{double,width=0.2mm,left=0.5}{v3,v6}
  \fmf{double,width=0.2mm,left=0.3}{v7,v9}
  \fmf{double,width=0.2mm,right=0.4}{v3,v8}
  \fmfv{decor.size=0, label=${\scriptstyle 1}$, l.dist=1mm, l.angle=-180}{v1}
  \fmfv{decor.size=0, label=${\scriptstyle 3}$, l.dist=1mm, l.angle=180}{v2}
  \fmfv{decor.size=0, label=${\scriptstyle 4}$, l.dist=1mm, l.angle=45}{v12}
  \fmfv{decor.size=0, label=${\scriptstyle 2}$, l.dist=1mm, l.angle=-45}{v13}
  \fmfdot{v3}
  \end{fmfgraph*} } } 
%%%%%%
\hspace*{3mm} + \hspace*{2mm} \frac{1}{6} \hspace*{2mm}
%%%%
  \parbox{14mm}{\centerline{
  \begin{fmfgraph*}(11,11)
  \setval
  \fmfforce{1/11w,0h}{v1}
  \fmfforce{0w,1h}{v2}
  \fmfforce{3/11w,2/11h}{v3}
  \fmfforce{2/11w,9/11h}{v4}
  \fmfforce{3/11w,6.5/11h}{v5}
  \fmfforce{1.5/11w,7.5/11h}{v6}
  \fmfforce{4.5/11w,8.5/11h}{v7}
  \fmfforce{7/11w,4.5/11h}{v8}
  \fmfforce{7/11w,6.5/11h}{v9}
  \fmfforce{9.2/11w,6.7/11h}{v10}
  \fmfforce{9/11w,4.5/11h}{v11}
  \fmfforce{1w,8.5/11h}{v12}
  \fmfforce{1w,2.5/11h}{v13}
  \fmf{plain}{v1,v3}
  \fmf{plain}{v2,v4}
  \fmf{plain}{v10,v12}
  \fmf{plain}{v11,v13}
  \fmf{double,width=0.2mm,left=1}{v6,v7,v6}
  \fmf{double,width=0.2mm,left=1}{v9,v11,v9}
  \fmf{double,width=0.2mm}{v3,v5}
  \fmf{double,width=0.2mm,left=0.5}{v3,v6}
  \fmf{double,width=0.2mm,left=0.3}{v7,v9}
  \fmf{double,width=0.2mm,right=0.4}{v3,v8}
  \fmfv{decor.size=0, label=${\scriptstyle 1}$, l.dist=1mm, l.angle=-180}{v1}
  \fmfv{decor.size=0, label=${\scriptstyle 4}$, l.dist=1mm, l.angle=180}{v2}
  \fmfv{decor.size=0, label=${\scriptstyle 2}$, l.dist=1mm, l.angle=45}{v12}
  \fmfv{decor.size=0, label=${\scriptstyle 3}$, l.dist=1mm, l.angle=-45}{v13}
  \fmfdot{v3}
  \end{fmfgraph*} } }  \hspace*{0.2cm}\, .
\label{PELSGAM}
\end{eqnarray}
Note that Eqs. (\ref{PELSSELF}), (\ref{PELSDY}) and (\ref{PELSGAM}) only generate 
one-particle irreducible diagrams\index{one-particle irreducible diagrams} 
for $\Sigma$ and $\Gamma$, if all lower
loop orders contain one-particle irreducible 
diagrams\index{one-particle irreducible diagrams}. 
By induction this establishes 
that $\Sigma$ and $\Gamma$ only consist of
one-particle irreducible diagrams\index{one-particle irreducible diagrams}.
\section{Tadpole\index{tadpole}-Free 
One-Particle Irreducible Diagrams\index{one-particle irreducible diagrams}} 
\label{PELSTadpole}
In order to reduce the number of diagrams, we aim at
substituting the free correlation function
$G$ by a modified one $\tilde{\mbox{\boldmath $G$}}$. If we would
fix $\tilde{\mbox{\boldmath $G$}}$ according to
\begin{eqnarray} 
\label{PELSVACC}
\tilde{\mbox{\boldmath $G$}}_{12}^{-1} = G^{-1}_{12}
+ \frac{1}{2} \int_{34} V_{1234} \tilde{\mbox{\boldmath $G$}}_{34}
\, ,
\end{eqnarray}
it would contain all repetitive one-loop corrections. This method was 
established
in Ref.\cite{PELSBoris2} to get rid of all one-particle
irreducible diagrams\index{one-particle irreducible diagrams} 
carrying tadpole\index{tadpole} 
corrections when calculating the
$\beta$-function of the vacuum energy density up to five loops.
Here, however, we go one step further by demanding instead of 
(\ref{PELSVACC})
\begin{eqnarray} 
\label{PELSMODIFY}
\tilde{\mbox{\boldmath $G$}}_{12}^{-1} = G^{-1}_{12}
+ \frac{1}{2} \int_{34} V_{1234} \mbox{\boldmath $G$}_{34}
\, ,
\end{eqnarray}
which amounts to absorbing all momentum-independent line corrections into 
the mass\cite{PELSMuenster}. The tadpole\index{tadpole} 
corrections of the modified
correlation function $\tilde{\mbox{\boldmath $G$}}$ arising from
(\ref{PELSMODIFY}) were already treated perturbatively in Ref.\cite{PELSBoris1}.
It was shown that they lead to additional diagrams which cancel order by 
order all diagrams of one-particle irreducible 
$n$-point functions\index{one-particle irreducible $n$-point function} 
carrying any kind of 
tadpole\index{tadpole}
correction. In Ref.\cite{PELSKonstantin} we elaborate that these inefficient
cancellations can be circumvented as the remaining 
tadpole-free\index{tadpole}
one-particle irreducible diagrams\index{one-particle irreducible diagrams} 
directly follow from a 
closed set of functional 
differential equations\index{functional differential equation}. 
To this end we consider
the modified self-energy\index{self-energy}
\begin{eqnarray}
\label{PELSMSE} 
\tilde{\Sigma}_{12} = 
\tilde{\mbox{\boldmath $G$}}_{12}^{-1}
-\mbox{\boldmath $G$}_{12}^{-1}\, 
\end{eqnarray}
and the one-particle irreducible four-point 
function\index{one-particle irreducible four-point function}
$\Gamma$
as functionals of the modified correlation function
$\tilde{\mbox{\boldmath $G$}}$.
Extending the elements of 
Feynman diagrams\index{Feynman diagrams} 
by a wiggly line representing the modified correlation 
function
\begin{eqnarray}
\label{PELSMDC}
\parbox{10mm}{\centerline{
\begin{fmfgraph*}(7,3)
\setval
\fmfforce{0w,1/2h}{v1}
\fmfforce{1w,1/2h}{v2}
\fmf{wiggly,width=0.2mm}{v2,v1}
\fmfv{decor,size=0, label=${\scriptstyle 1}$, l.dist=1mm, l.angle=-180}{v1}
\fmfv{decor,size=0, label=${\scriptstyle 2}$, l.dist=1mm, l.angle=0}{v2}
\end{fmfgraph*} } } 
\, = \, \tilde{\mbox{\boldmath $G$}}_{12} 
\end{eqnarray}
and by a $2$-vertex with a wiggly dot representing the 
tadpole-free\index{tadpole}
self-energy\index{self-energy}
\begin{eqnarray}
\parbox{11mm}{\centerline{
\begin{fmfgraph*}(9,3)
\setval
\fmfforce{0w,1/2h}{v1}
\fmfforce{3/9w,1/2h}{v2}
\fmfforce{6/9w,1/2h}{v3}
\fmfforce{1w,1/2h}{v4}
\fmfforce{1/2w,1h}{v5}
\fmfforce{1/2w,0h}{v6}
\fmf{wiggly,width=0.2mm}{v1,v2}
\fmf{wiggly,width=0.2mm}{v3,v4}
\fmf{dbl_wiggly,width=0.2mm,left=1}{v5,v6,v5}
\fmfv{decor.size=0, label=${\scriptstyle 1}$, l.dist=1mm, l.angle=-180}{v1}
\fmfv{decor.size=0, label=${\scriptstyle 2}$, l.dist=1mm, l.angle=0}{v4}
\end{fmfgraph*} } } 
\, = \, \tilde{\Sigma}_{12} \hspace*{0.3cm}
\end{eqnarray}
our results read as follows. The tadpole-free\index{tadpole} 
self-energy\index{self-energy} $\tilde{\Sigma}$
obeys the integral equation
\begin{eqnarray}
\label{PELSWIG1}
\parbox{11mm}{\centerline{
\begin{fmfgraph*}(9,3)
\setval
\fmfforce{0w,1/2h}{v1}
\fmfforce{3/9w,1/2h}{v2}
\fmfforce{6/9w,1/2h}{v3}
\fmfforce{1w,1/2h}{v4}
\fmfforce{1/2w,1h}{v5}
\fmfforce{1/2w,0h}{v6}
\fmf{wiggly,width=0.2mm}{v1,v2}
\fmf{wiggly,width=0.2mm}{v3,v4}
\fmf{dbl_wiggly,width=0.2mm,left=1}{v5,v6,v5}
\fmfv{decor.size=0, label=${\scriptstyle 1}$, l.dist=1mm, l.angle=-180}{v1}
\fmfv{decor.size=0, label=${\scriptstyle 2}$, l.dist=1mm, l.angle=0}{v4}
\end{fmfgraph*} } } 
\hspace*{5mm} = \hspace*{3mm} \frac{1}{6} \hspace*{3mm}
\parbox{17mm}{\centerline{
\begin{fmfgraph*}(14,6)
\setval
\fmfforce{0w,1/2h}{v1}
\fmfforce{3/14w,1/2h}{v2}
\fmfforce{8/14w,1/2h}{v3}
\fmfforce{11/14w,1/2h}{v4}
\fmfforce{1w,1/2h}{v5}
\fmfforce{9.2/14w,2.2/3h}{v6}
\fmfforce{9.2/14w,0.8/3h}{v7}
\fmf{wiggly,width=0.2mm}{v1,v2}
\fmf{wiggly,width=0.2mm}{v4,v5}
\fmf{double,width=0.2mm,left=0.8}{v2,v6}
\fmf{double,width=0.2mm}{v2,v3}
\fmf{double,width=0.2mm,right=0.8}{v2,v7}
\fmf{double,width=0.2mm,right=1}{v3,v4,v3}
\fmfv{decor.size=0, label=${\scriptstyle 1}$, l.dist=1mm, l.angle=-180}{v1}
\fmfv{decor.size=0, label=${\scriptstyle 2}$, l.dist=1mm, l.angle=0}{v5}
\fmfdot{v2}
\end{fmfgraph*} } } \hspace*{0.2cm},
\end{eqnarray}
where the fully-interacting 
two-point function $\mbox{\boldmath $G$}^{{\rm c}}$
is obtained from (\ref{PELSMSE}) according to the modified Dyson-equation
\begin{eqnarray}
\label{PELSWIG2}
\parbox{10mm}{\centerline{
\begin{fmfgraph*}(7,3)
\setval
\fmfforce{0w,1/2h}{v1}
\fmfforce{1w,1/2h}{v2}
\fmf{double,width=0.2mm}{v2,v1}
\fmfv{decor,size=0, label=${\scriptstyle 1}$, l.dist=1mm, l.angle=-180}{v1}
\fmfv{decor,size=0, label=${\scriptstyle 2}$, l.dist=1mm, l.angle=0}{v2}
\end{fmfgraph*} } } 
\hspace*{3mm} = \hspace*{3mm}
\parbox{10mm}{\centerline{
\begin{fmfgraph*}(7,3)
\setval
\fmfforce{0w,1/2h}{v1}
\fmfforce{1w,1/2h}{v2}
\fmf{wiggly,width=0.2mm}{v2,v1}
\fmfv{decor,size=0, label=${\scriptstyle 1}$, l.dist=1mm, l.angle=-180}{v1}
\fmfv{decor,size=0, label=${\scriptstyle 2}$, l.dist=1mm, l.angle=0}{v2}
\end{fmfgraph*} } } 
\hspace*{3mm} + \hspace*{3mm}
\parbox{15mm}{\centerline{
\begin{fmfgraph*}(13,3)
\setval
\fmfforce{0w,1/2h}{v1}
\fmfforce{5/13w,1/2h}{v2}
\fmfforce{8/13w,1/2h}{v3}
\fmfforce{1w,1/2h}{v4}
\fmfforce{1/2w,1h}{v5}
\fmfforce{1/2w,0h}{v6}
\fmf{wiggly,width=0.2mm}{v1,v2}
\fmf{double,width=0.2mm}{v3,v4}
\fmf{dbl_wiggly,width=0.2mm,left=1}{v5,v6,v5}
\fmfv{decor.size=0, label=${\scriptstyle 1}$, l.dist=1mm, l.angle=-180}{v1}
\fmfv{decor.size=0, label=${\scriptstyle 2}$, l.dist=1mm, l.angle=0}{v4}
\end{fmfgraph*} } } \hspace*{0.2cm}.
\end{eqnarray}
Then the one-particle irreducible four-point 
function\index{one-particle irreducible four-point function} $\Gamma$
obeys \vspace*{0.3cm}
\begin{eqnarray}
  \parbox{10mm}{\centerline{
  \begin{fmfgraph*}(6,6)
  \setval
  \fmfforce{0w,0h}{v1}
  \fmfforce{0w,1h}{v2}
  \fmfforce{1w,1h}{v3}
  \fmfforce{1w,0h}{v4}
  \fmfforce{1/3w,1/3h}{v5}
  \fmfforce{1/3w,2/3h}{v6}
  \fmfforce{2/3w,2/3h}{v7}
  \fmfforce{2/3w,1/3h}{v8}
  \fmf{wiggly,width=0.2mm}{v1,v5}
  \fmf{wiggly,width=0.2mm}{v2,v6}
  \fmf{wiggly,width=0.2mm}{v3,v7}
  \fmf{wiggly,width=0.2mm}{v4,v8}
  \fmf{double,width=0.2mm,left=1}{v5,v7,v5}
  \fmfv{decor.size=0, label=${\scriptstyle 1}$, l.dist=1mm, l.angle=-135}{v1}
  \fmfv{decor.size=0, label=${\scriptstyle 2}$, l.dist=1mm, l.angle=135}{v2}
  \fmfv{decor.size=0, label=${\scriptstyle 3}$, l.dist=1mm, l.angle=45}{v3}
  \fmfv{decor.size=0, label=${\scriptstyle 4}$, l.dist=1mm, l.angle=-45}{v4}
  \end{fmfgraph*} } } 
%%%%
\hspace*{3mm} & = & \hspace*{3mm}
%%%%
  \parbox{7mm}{\centerline{
  \begin{fmfgraph*}(4.5,4.5)
  \setval
  \fmfforce{0w,0h}{v1}
  \fmfforce{0w,1h}{v2}
  \fmfforce{1w,1h}{v3}
  \fmfforce{1w,0h}{v4}
  \fmfforce{1/2w,1/2h}{v5}
  \fmf{wiggly,width=0.2mm}{v1,v3}
  \fmf{wiggly,width=0.2mm}{v2,v4}
  \fmfv{decor.size=0, label=${\scriptstyle 1}$, l.dist=1mm, l.angle=-135}{v1}
  \fmfv{decor.size=0, label=${\scriptstyle 2}$, l.dist=1mm, l.angle=135}{v2}
  \fmfv{decor.size=0, label=${\scriptstyle 3}$, l.dist=1mm, l.angle=45}{v3}
  \fmfv{decor.size=0, label=${\scriptstyle 4}$, l.dist=1mm, l.angle=-45}{v4}
  \fmfdot{v5}
  \end{fmfgraph*} } } 
%%%%
\hspace*{3mm} + \hspace*{2mm} \frac{1}{3} \hspace*{3mm}
%%%%
  \parbox{10mm}{\centerline{
  \begin{fmfgraph*}(8,8)
  \setval
  \fmfforce{0w,5.5/8h}{v1}
  \fmfforce{3/8w,5.5/8h}{v2}
  \fmfforce{1w,5.5/8h}{v3}
  \fmfforce{1w,1.5/8h}{v4}
  \fmfforce{1w,9.5/8h}{v5}
  \fmf{wiggly,width=0.2mm}{v1,v2}
  \fmf{wiggly,width=0.2mm}{v2,v3}
  \fmf{wiggly,width=0.2mm,right=0.3}{v2,v4}
  \fmf{double,width=0.2mm,left=0.3}{v2,v5}
  \fmfv{decor.size=0, label=${\scriptstyle 1}$, l.dist=1mm, l.angle=-180}{v1}
  \fmfv{decor.size=0, label=${\scriptstyle 6}$, l.dist=1mm, l.angle=0}{v3}
  \fmfv{decor.size=0, label=${\scriptstyle 7}$, l.dist=1mm, l.angle=0}{v4}
  \fmfv{decor.size=0, label=${\scriptstyle 5}$, l.dist=1mm, l.angle=0}{v5}
  \fmfdot{v2}
  \end{fmfgraph*} } } 
%%%%
\hspace*{3mm} 
%%%%
\dphiwiggly{
  \parbox{11mm}{\centerline{
  \begin{fmfgraph*}(6,7)
  \setval
  \fmfforce{0w,2/7h}{v1}
  \fmfforce{0w,8/7h}{v2}
  \fmfforce{1w,8/7h}{v3}
  \fmfforce{1w,2/7h}{v4}
  \fmfforce{1/3w,4/7h}{v5}
  \fmfforce{1/3w,6/7h}{v6}
  \fmfforce{2/3w,6/7h}{v7}
  \fmfforce{2/3w,4/7h}{v8}
  \fmf{wiggly,width=0.2mm}{v1,v5}
  \fmf{wiggly,width=0.2mm}{v2,v6}
  \fmf{wiggly,width=0.2mm}{v3,v7}
  \fmf{wiggly,width=0.2mm}{v4,v8}
  \fmf{double,width=0.2mm,left=1}{v5,v7,v5}
  \fmfv{decor.size=0, label=${\scriptstyle 5}$, l.dist=0.5mm, l.angle=-180}{v1}
  \fmfv{decor.size=0, label=${\scriptstyle 2}$, l.dist=0.5mm, l.angle=180}{v2}
  \fmfv{decor.size=0, label=${\scriptstyle 3}$, l.dist=0.5mm, l.angle=0}{v3}
  \fmfv{decor.size=0, label=${\scriptstyle 4}$, l.dist=0.5mm, l.angle=0}{v4}
  \end{fmfgraph*} } } 
}{6}{7}
%%%%
%\nonumber \\ && \nonumber \\ && 
\hspace*{0.2cm} + \hspace*{2mm} \frac{1}{6} \hspace*{1mm}
%%%%
  \parbox{10mm}{\centerline{
  \begin{fmfgraph*}(7,8)
  \setval
  \fmfforce{0w,9.5/8h}{v1}
  \fmfforce{2/7w,7.5/8h}{v2}
  \fmfforce{1w,9.5/8h}{v3}
  \fmfforce{2/7w,3.5/8h}{v4}
  \fmfforce{1w,4.5/8h}{v5}
  \fmfforce{1w,1.5/8h}{v6}
  \fmf{wiggly,width=0.2mm}{v1,v2}
  \fmf{double,width=0.2mm,left=0.1}{v2,v3}
  \fmf{double,width=0.2mm,left=0.5}{v2,v4}
  \fmf{double,width=0.2mm,right=0.5}{v2,v4}
  \fmf{wiggly,width=0.2mm}{v4,v5}
  \fmf{wiggly,width=0.2mm,right=0.2}{v4,v6}
  \fmfv{decor.size=0, label=${\scriptstyle 1}$, l.dist=0.5mm, l.angle=135}{v1}
  \fmfv{decor.size=0, label=${\scriptstyle 5}$, l.dist=1mm, l.angle=0}{v3}
  \fmfv{decor.size=0, label=${\scriptstyle 6}$, l.dist=1mm, l.angle=25}{v5}
  \fmfv{decor.size=0, label=${\scriptstyle 7}$, l.dist=1mm, l.angle=0}{v6}
  \fmfdot{v2}
  \fmfdot{v4}
  \end{fmfgraph*} } } 
%%%%
\hspace*{3mm} 
%%%%
\dphiwiggly{
  \parbox{11mm}{\centerline{
  \begin{fmfgraph*}(6,7)
  \setval
  \fmfforce{0w,2/7h}{v1}
  \fmfforce{0w,8/7h}{v2}
  \fmfforce{1w,8/7h}{v3}
  \fmfforce{1w,2/7h}{v4}
  \fmfforce{1/3w,4/7h}{v5}
  \fmfforce{1/3w,6/7h}{v6}
  \fmfforce{2/3w,6/7h}{v7}
  \fmfforce{2/3w,4/7h}{v8}
  \fmf{wiggly,width=0.2mm}{v1,v5}
  \fmf{wiggly,width=0.2mm}{v2,v6}
  \fmf{wiggly,width=0.2mm}{v3,v7}
  \fmf{wiggly,width=0.2mm}{v4,v8}
  \fmf{double,width=0.2mm,left=1}{v5,v7,v5}
  \fmfv{decor.size=0, label=${\scriptstyle 5}$, l.dist=0.5mm, l.angle=-180}{v1}
  \fmfv{decor.size=0, label=${\scriptstyle 2}$, l.dist=0.5mm, l.angle=180}{v2}
  \fmfv{decor.size=0, label=${\scriptstyle 3}$, l.dist=0.5mm, l.angle=0}{v3}
  \fmfv{decor.size=0, label=${\scriptstyle 4}$, l.dist=0.5mm, l.angle=0}{v4}
  \end{fmfgraph*} } } 
}{6}{7}
%%%%
\hspace*{3mm} + \hspace*{2mm} \frac{1}{12} \hspace*{1.5mm}
%%%%
  \parbox{12mm}{\centerline{
  \begin{fmfgraph*}(8,12)
  \setval
  \fmfforce{2/8w,14.5/12h}{v1}
  \fmfforce{4/8w,12.5/12h}{v2}
  \fmfforce{1w,13.5/12h}{v3}
  \fmfforce{4/8w,4.5/12h}{v4}
  \fmfforce{1.5/8w,10/12h}{v5}
  \fmfforce{1.5/8w,7/12h}{v6}
  \fmfforce{2.7/8w,9/12h}{v7}
  \fmfforce{2.7/8w,8/12h}{v8}
  \fmfforce{1w,6.5/12h}{v9}
  \fmfforce{1w,3.5/12h}{v10}
  \fmf{wiggly,width=0.2mm}{v1,v2}
  \fmf{double,width=0.2mm,left=0.1}{v2,v3}
  \fmf{double,width=0.2mm,right=0.25}{v2,v5}
  \fmf{double,width=0.2mm,left=0.35}{v2,v7}
  \fmf{double,width=0.2mm,right=0.25}{v6,v4}
  \fmf{double,width=0.2mm,left=0.35}{v8,v4}
  \fmf{double,width=0.2mm,left=1}{v5,v6,v5}
  \fmf{wiggly,width=0.2mm}{v4,v9}
  \fmf{wiggly,width=0.2mm,right=0.1}{v4,v10}
  \fmfv{decor.size=0, label=${\scriptstyle 1}$, l.dist=0.5mm, l.angle=165}{v1}
  \fmfv{decor.size=0, label=${\scriptstyle 5}$, l.dist=1mm, l.angle=0}{v3}
  \fmfv{decor.size=0, label=${\scriptstyle 6}$, l.dist=1mm, l.angle=25}{v9}
  \fmfv{decor.size=0, label=${\scriptstyle 7}$, l.dist=1mm, l.angle=0}{v10}
  \fmfdot{v2}
  \fmfdot{v4}
  \end{fmfgraph*} } } 
%%%%
\hspace*{3mm} 
%%%%
\dphiwiggly{
  \parbox{11mm}{\centerline{
  \begin{fmfgraph*}(6,7)
  \setval
  \fmfforce{0w,2/7h}{v1}
  \fmfforce{0w,8/7h}{v2}
  \fmfforce{1w,8/7h}{v3}
  \fmfforce{1w,2/7h}{v4}
  \fmfforce{1/3w,4/7h}{v5}
  \fmfforce{1/3w,6/7h}{v6}
  \fmfforce{2/3w,6/7h}{v7}
  \fmfforce{2/3w,4/7h}{v8}
  \fmf{wiggly,width=0.2mm}{v1,v5}
  \fmf{wiggly,width=0.2mm}{v2,v6}
  \fmf{wiggly,width=0.2mm}{v3,v7}
  \fmf{wiggly,width=0.2mm}{v4,v8}
  \fmf{double,width=0.2mm,left=1}{v5,v7,v5}
  \fmfv{decor.size=0, label=${\scriptstyle 5}$, l.dist=0.5mm, l.angle=-180}{v1}
  \fmfv{decor.size=0, label=${\scriptstyle 2}$, l.dist=0.5mm, l.angle=180}{v2}
  \fmfv{decor.size=0, label=${\scriptstyle 3}$, l.dist=0.5mm, l.angle=0}{v3}
  \fmfv{decor.size=0, label=${\scriptstyle 4}$, l.dist=0.5mm, l.angle=0}{v4}
  \end{fmfgraph*} } } 
}{6}{7}
%%%%
%%%%
\hspace*{0.2cm} + \hspace*{2mm} \frac{1}{2} \hspace*{2mm}
%%%%                     %%%%%%%%%%%%%%%%%%
  \parbox{14mm}{\centerline{
  \begin{fmfgraph*}(11,6)
  \setval
  \fmfforce{0w,1/6h}{v1}
  \fmfforce{0w,5/6h}{v2}
  \fmfforce{1w,1h}{v3}
  \fmfforce{1w,0h}{v4}
  \fmfforce{7/11w,1/3h}{v5}
  \fmfforce{7/11w,2/3h}{v6}
  \fmfforce{9/11w,2/3h}{v7}
  \fmfforce{9/11w,1/3h}{v8}
  \fmfforce{2/11w,1/2h}{v9}
  \fmf{wiggly,width=0.2mm}{v1,v9}
  \fmf{wiggly,width=0.2mm}{v2,v9}
  \fmf{wiggly,width=0.2mm}{v3,v7}
  \fmf{wiggly,width=0.2mm}{v4,v8}
  \fmf{double,width=0.2mm,left=1}{v5,v7,v5}
  \fmf{double,width=0.2mm,left=0.7}{v9,v6}
  \fmf{double,width=0.2mm,right=0.7}{v9,v5}
  \fmfv{decor.size=0, label=${\scriptstyle 1}$, l.dist=1mm, l.angle=-135}{v1}
  \fmfv{decor.size=0, label=${\scriptstyle 2}$, l.dist=1mm, l.angle=135}{v2}
  \fmfv{decor.size=0, label=${\scriptstyle 3}$, l.dist=1mm, l.angle=25}{v3}
  \fmfv{decor.size=0, label=${\scriptstyle 4}$, l.dist=1mm, l.angle=-25}{v4}
  \fmfdot{v9}
  \end{fmfgraph*} } } 
%%%%
\nonumber \\ && \nonumber \\
%%%%
& & 
+ \hspace*{2mm} \frac{1}{2} \hspace*{2mm}
%%%%
  \parbox{14mm}{\centerline{
  \begin{fmfgraph*}(11,6)
  \setval
  \fmfforce{0w,1/6h}{v1}
  \fmfforce{0w,5/6h}{v2}
  \fmfforce{1w,1h}{v3}
  \fmfforce{1w,0h}{v4}
  \fmfforce{7/11w,1/3h}{v5}
  \fmfforce{7/11w,2/3h}{v6}
  \fmfforce{9/11w,2/3h}{v7}
  \fmfforce{9/11w,1/3h}{v8}
  \fmfforce{2/11w,1/2h}{v9}
  \fmf{wiggly,width=0.2mm}{v1,v9}
  \fmf{wiggly,width=0.2mm}{v2,v9}
  \fmf{wiggly,width=0.2mm}{v3,v7}
  \fmf{wiggly,width=0.2mm}{v4,v8}
  \fmf{double,width=0.2mm,left=1}{v5,v7,v5}
  \fmf{double,width=0.2mm,left=0.7}{v9,v6}
  \fmf{double,width=0.2mm,right=0.7}{v9,v5}
  \fmfv{decor.size=0, label=${\scriptstyle 1}$, l.dist=1mm, l.angle=-135}{v1}
  \fmfv{decor.size=0, label=${\scriptstyle 3}$, l.dist=1mm, l.angle=135}{v2}
  \fmfv{decor.size=0, label=${\scriptstyle 4}$, l.dist=1mm, l.angle=25}{v3}
  \fmfv{decor.size=0, label=${\scriptstyle 2}$, l.dist=1mm, l.angle=-25}{v4}
  \fmfdot{v9}
  \end{fmfgraph*} } } 
%%%%
\hspace*{3mm} + \hspace*{2mm} \frac{1}{2} \hspace*{2mm}
%%%%
  \parbox{14mm}{\centerline{
  \begin{fmfgraph*}(11,6)
  \setval
  \fmfforce{0w,1/6h}{v1}
  \fmfforce{0w,5/6h}{v2}
  \fmfforce{1w,1h}{v3}
  \fmfforce{1w,0h}{v4}
  \fmfforce{7/11w,1/3h}{v5}
  \fmfforce{7/11w,2/3h}{v6}
  \fmfforce{9/11w,2/3h}{v7}
  \fmfforce{9/11w,1/3h}{v8}
  \fmfforce{2/11w,1/2h}{v9}
  \fmf{wiggly,width=0.2mm}{v1,v9}
  \fmf{wiggly,width=0.2mm}{v2,v9}
  \fmf{wiggly,width=0.2mm}{v3,v7}
  \fmf{wiggly,width=0.2mm}{v4,v8}
  \fmf{double,width=0.2mm,left=1}{v5,v7,v5}
  \fmf{double,width=0.2mm,left=0.7}{v9,v6}
  \fmf{double,width=0.2mm,right=0.7}{v9,v5}
  \fmfv{decor.size=0, label=${\scriptstyle 1}$, l.dist=1mm, l.angle=-135}{v1}
  \fmfv{decor.size=0, label=${\scriptstyle 4}$, l.dist=1mm, l.angle=135}{v2}
  \fmfv{decor.size=0, label=${\scriptstyle 2}$, l.dist=1mm, l.angle=25}{v3}
  \fmfv{decor.size=0, label=${\scriptstyle 3}$, l.dist=1mm, l.angle=-25}{v4}
  \fmfdot{v9}
  \end{fmfgraph*} } } 
%%%%
%%%%
%\nonumber \\ && \nonumber \\
%%%%
%& & 
\hspace*{0.2cm} + \hspace*{2mm} \frac{1}{6} \hspace*{2mm}
%%%%                  %%%%%%%%%%%%%%%%%%%%%%
  \parbox{14mm}{\centerline{
  \begin{fmfgraph*}(11,11)
  \setval
  \fmfforce{1/11w,0h}{v1}
  \fmfforce{0w,1h}{v2}
  \fmfforce{3/11w,2/11h}{v3}
  \fmfforce{2/11w,9/11h}{v4}
  \fmfforce{3/11w,6.5/11h}{v5}
  \fmfforce{1.5/11w,7.5/11h}{v6}
  \fmfforce{4.5/11w,8.5/11h}{v7}
  \fmfforce{7/11w,4.5/11h}{v8}
  \fmfforce{7/11w,6.5/11h}{v9}
  \fmfforce{9.2/11w,6.7/11h}{v10}
  \fmfforce{9/11w,4.5/11h}{v11}
  \fmfforce{1w,8.5/11h}{v12}
  \fmfforce{1w,2.5/11h}{v13}
  \fmf{wiggly,width=0.2mm}{v1,v3}
  \fmf{wiggly,width=0.2mm}{v2,v4}
  \fmf{wiggly,width=0.2mm}{v10,v12}
  \fmf{wiggly,width=0.2mm}{v11,v13}
  \fmf{double,width=0.2mm,left=1}{v6,v7,v6}
  \fmf{double,width=0.2mm,left=1}{v9,v11,v9}
  \fmf{double,width=0.2mm}{v3,v5}
  \fmf{double,width=0.2mm,left=0.5}{v3,v6}
  \fmf{double,width=0.2mm,left=0.3}{v7,v9}
  \fmf{double,width=0.2mm,right=0.4}{v3,v8}
  \fmfv{decor.size=0, label=${\scriptstyle 1}$, l.dist=1mm, l.angle=-180}{v1}
  \fmfv{decor.size=0, label=${\scriptstyle 2}$, l.dist=1mm, l.angle=180}{v2}
  \fmfv{decor.size=0, label=${\scriptstyle 3}$, l.dist=1mm, l.angle=45}{v12}
  \fmfv{decor.size=0, label=${\scriptstyle 4}$, l.dist=1mm, l.angle=-45}{v13}
  \fmfdot{v3}
  \end{fmfgraph*} } } 
%%%%%
\hspace*{3mm} + \hspace*{2mm} \frac{1}{6} \hspace*{2mm}
%%%%
  \parbox{14mm}{\centerline{
  \begin{fmfgraph*}(11,11)
  \setval
  \fmfforce{1/11w,0h}{v1}
  \fmfforce{0w,1h}{v2}
  \fmfforce{3/11w,2/11h}{v3}
  \fmfforce{2/11w,9/11h}{v4}
  \fmfforce{3/11w,6.5/11h}{v5}
  \fmfforce{1.5/11w,7.5/11h}{v6}
  \fmfforce{4.5/11w,8.5/11h}{v7}
  \fmfforce{7/11w,4.5/11h}{v8}
  \fmfforce{7/11w,6.5/11h}{v9}
  \fmfforce{9.2/11w,6.7/11h}{v10}
  \fmfforce{9/11w,4.5/11h}{v11}
  \fmfforce{1w,8.5/11h}{v12}
  \fmfforce{1w,2.5/11h}{v13}
  \fmf{wiggly,width=0.2mm}{v1,v3}
  \fmf{wiggly,width=0.2mm}{v2,v4}
  \fmf{wiggly,width=0.2mm}{v10,v12}
  \fmf{wiggly,width=0.2mm}{v11,v13}
  \fmf{double,width=0.2mm,left=1}{v6,v7,v6}
  \fmf{double,width=0.2mm,left=1}{v9,v11,v9}
  \fmf{double,width=0.2mm}{v3,v5}
  \fmf{double,width=0.2mm,left=0.5}{v3,v6}
  \fmf{double,width=0.2mm,left=0.3}{v7,v9}
  \fmf{double,width=0.2mm,right=0.4}{v3,v8}
  \fmfv{decor.size=0, label=${\scriptstyle 1}$, l.dist=1mm, l.angle=-180}{v1}
  \fmfv{decor.size=0, label=${\scriptstyle 3}$, l.dist=1mm, l.angle=180}{v2}
  \fmfv{decor.size=0, label=${\scriptstyle 4}$, l.dist=1mm, l.angle=45}{v12}
  \fmfv{decor.size=0, label=${\scriptstyle 2}$, l.dist=1mm, l.angle=-45}{v13}
  \fmfdot{v3}
  \end{fmfgraph*} } } 
%%%%%%
\hspace*{3mm} + \hspace*{2mm} \frac{1}{6} \hspace*{2mm}
%%%%
  \parbox{14mm}{\centerline{
  \begin{fmfgraph*}(11,11)
  \setval
  \fmfforce{1/11w,0h}{v1}
  \fmfforce{0w,1h}{v2}
  \fmfforce{3/11w,2/11h}{v3}
  \fmfforce{2/11w,9/11h}{v4}
  \fmfforce{3/11w,6.5/11h}{v5}
  \fmfforce{1.5/11w,7.5/11h}{v6}
  \fmfforce{4.5/11w,8.5/11h}{v7}
  \fmfforce{7/11w,4.5/11h}{v8}
  \fmfforce{7/11w,6.5/11h}{v9}
  \fmfforce{9.2/11w,6.7/11h}{v10}
  \fmfforce{9/11w,4.5/11h}{v11}
  \fmfforce{1w,8.5/11h}{v12}
  \fmfforce{1w,2.5/11h}{v13}
  \fmf{wiggly,width=0.2mm}{v1,v3}
  \fmf{wiggly,width=0.2mm}{v2,v4}
  \fmf{wiggly,width=0.2mm}{v10,v12}
  \fmf{wiggly,width=0.2mm}{v11,v13}
  \fmf{double,width=0.2mm,left=1}{v6,v7,v6}
  \fmf{double,width=0.2mm,left=1}{v9,v11,v9}
  \fmf{double,width=0.2mm}{v3,v5}
  \fmf{double,width=0.2mm,left=0.5}{v3,v6}
  \fmf{double,width=0.2mm,left=0.3}{v7,v9}
  \fmf{double,width=0.2mm,right=0.4}{v3,v8}
  \fmfv{decor.size=0, label=${\scriptstyle 1}$, l.dist=1mm, l.angle=-180}{v1}
  \fmfv{decor.size=0, label=${\scriptstyle 4}$, l.dist=1mm, l.angle=180}{v2}
  \fmfv{decor.size=0, label=${\scriptstyle 2}$, l.dist=1mm, l.angle=45}{v12}
  \fmfv{decor.size=0, label=${\scriptstyle 3}$, l.dist=1mm, l.angle=-45}{v13}
  \fmfdot{v3}
  \end{fmfgraph*} } } \hspace*{0.2cm} . 
\label{PELSWIG3}
\end{eqnarray}\\
Iteratively solving Eqs.~(\ref{PELSWIG1})--(\ref{PELSWIG3}) in a graphical way
leads to all tadpole-free\index{tadpole} diagrams of the 
self-energy\index{self-energy} and of the 
one-particle irreducible four-point 
function\index{one-particle irreducible four-point function}
 together with their weights\index{weights}. 
Up to five loops they are
listed in Appendix A of Ref.\cite{PELSVerena}. Their respective quadratic
and logarithmic divergencies in $d=4-\epsilon$ dimensions
contribute to the $1/\epsilon$-poles of the 
renormalization\index{renormalization} constants
of the field
$\phi$, the coupling constant $g$ and the mass $m^2$ within the minimal
subtraction scheme, so that they determine the critical 
exponents\index{critical exponents} of
scalar $\phi^4$-theory\index{$\phi^4$-theory}.
\section{One-Particle Irreducible Diagrams\index{one-particle irreducible diagrams} Without Line Corrections}
\label{PELSLine}
The number of diagrams can be even further reduced by substituting the free
correlation function $G$ by the fully-interacting two-point function
$\mbox{\boldmath $G$}$ itself. Such a substitution was investigated 
in Ref.\cite{PELSKleinert1} 
within the framework of higher functional Legendre
transformations\index{Legendre transformation}. 
By considering the self-energy\index{self-energy} $\Sigma$
and the one-particle irreducible four-point 
function\index{one-particle irreducible four-point function} $\Gamma$
as functionals of the fully-interacting
two-point function $\mbox{\boldmath $G$}$, it can be shown that
they obey\cite{PELSKleinert1,PELSKonstantin} \vspace*{0.2cm}
\begin{eqnarray}
\label{PELSFull1}
\parbox{11mm}{\centerline{
\begin{fmfgraph*}(9,3)
\setval
\fmfforce{0w,1/2h}{v1}
\fmfforce{3/9w,1/2h}{v2}
\fmfforce{6/9w,1/2h}{v3}
\fmfforce{1w,1/2h}{v4}
\fmfforce{1/2w,1h}{v5}
\fmfforce{1/2w,0h}{v6}
\fmf{double,width=0.2mm}{v1,v2}
\fmf{double,width=0.2mm}{v3,v4}
\fmf{double,width=0.2mm,left=1}{v5,v6,v5}
\fmfv{decor.size=0, label=${\scriptstyle 1}$, l.dist=1mm, l.angle=-180}{v1}
\fmfv{decor.size=0, label=${\scriptstyle 2}$, l.dist=1mm, l.angle=0}{v4}
\end{fmfgraph*} } } 
& \hspace*{0.2cm} = \hspace*{0.2cm} & \frac{1}{2} \hspace*{3mm}
\parbox{9mm}{\centerline{
\begin{fmfgraph*}(6,5)
\setval
\fmfforce{0w,0h}{v1}
\fmfforce{1/2w,0h}{v2}
\fmfforce{1w,0h}{v3}
\fmfforce{1/12w,1/2h}{v4}
\fmfforce{11/12w,1/2h}{v5}
\fmf{double,width=0.2mm}{v1,v3}
\fmf{double,width=0.2mm,left=1}{v4,v5,v4}
\fmfv{decor.size=0, label=${\scriptstyle 1}$, l.dist=1mm, l.angle=-180}{v1}
\fmfv{decor.size=0, label=${\scriptstyle 2}$, l.dist=1mm, l.angle=0}{v3}
\fmfdot{v2}
\end{fmfgraph*} } }
\hspace*{3mm} + \hspace*{2mm} \frac{1}{6} \hspace*{3mm}
\parbox{17mm}{\centerline{
\begin{fmfgraph*}(14,6)
\setval
\fmfforce{0w,1/2h}{v1}
\fmfforce{3/14w,1/2h}{v2}
\fmfforce{8/14w,1/2h}{v3}
\fmfforce{11/14w,1/2h}{v4}
\fmfforce{1w,1/2h}{v5}
\fmfforce{9.2/14w,2.2/3h}{v6}
\fmfforce{9.2/14w,0.8/3h}{v7}
\fmf{double,width=0.2mm}{v1,v2}
\fmf{double,width=0.2mm}{v4,v5}
\fmf{double,width=0.2mm,left=0.8}{v2,v6}
\fmf{double,width=0.2mm}{v2,v3}
\fmf{double,width=0.2mm,right=0.8}{v2,v7}
\fmf{double,width=0.2mm,right=1}{v3,v4,v3}
\fmfv{decor.size=0, label=${\scriptstyle 1}$, l.dist=1mm, l.angle=-180}{v1}
\fmfv{decor.size=0, label=${\scriptstyle 2}$, l.dist=1mm, l.angle=0}{v5}
\fmfdot{v2}
\end{fmfgraph*} } } \hspace*{0.2cm} , \\ && \nonumber \\
\label{PELSFull2}
\parbox{10mm}{\centerline{
\begin{fmfgraph*}(6,6)
\setval
\fmfforce{0w,0h}{v1}
\fmfforce{0w,1h}{v2}
\fmfforce{1w,1h}{v3}
\fmfforce{1w,0h}{v4}
\fmfforce{1/3w,1/3h}{v5}
\fmfforce{1/3w,2/3h}{v6}
\fmfforce{2/3w,2/3h}{v7}
\fmfforce{2/3w,1/3h}{v8}
\fmf{double,width=0.2mm}{v1,v5}
\fmf{double,width=0.2mm}{v2,v6}
\fmf{double,width=0.2mm}{v3,v7}
\fmf{double,width=0.2mm}{v4,v8}
\fmf{double,width=0.2mm,left=1}{v5,v7,v5}
\fmfv{decor.size=0, label=${\scriptstyle 1}$, l.dist=1mm, l.angle=-135}{v1}
\fmfv{decor.size=0, label=${\scriptstyle 2}$, l.dist=1mm, l.angle=135}{v2}
\fmfv{decor.size=0, label=${\scriptstyle 3}$, l.dist=1mm, l.angle=45}{v3}
\fmfv{decor.size=0, label=${\scriptstyle 4}$, l.dist=1mm, l.angle=-45}{v4}
\end{fmfgraph*} } } 
& = & 2 \hspace*{2mm}
\dphidouble{
\parbox{14mm}{\centerline{
\begin{fmfgraph*}(9,3)
\setval
\fmfforce{0w,3/4h}{v1}
\fmfforce{3/9w,3/4h}{v2}
\fmfforce{6/9w,3/4h}{v3}
\fmfforce{1w,3/4h}{v4}
\fmfforce{1/2w,5/4h}{v5}
\fmfforce{1/2w,1/4h}{v6}
\fmf{double,width=0.2mm}{v1,v2}
\fmf{double,width=0.2mm}{v3,v4}
\fmf{double,width=0.2mm,left=1}{v5,v6,v5}
\fmfv{decor.size=0, label=${\scriptstyle 1}$, l.dist=0.5mm, l.angle=-180}{v1}
\fmfv{decor.size=0, label=${\scriptstyle 2}$, l.dist=0.5mm, l.angle=0}{v4}
\end{fmfgraph*} } }
}{3}{4} 
\hspace*{3mm} + \hspace*{3mm} 
\parbox{10mm}{\centerline{
\begin{fmfgraph*}(8,6)
\setval
\fmfforce{0w,0h}{v1}
\fmfforce{0w,1h}{v2}
\fmfforce{1w,1h}{v3}
\fmfforce{1w,0h}{v4}
\fmfforce{1/4w,1/3h}{v5}
\fmfforce{1/4w,2/3h}{v6}
\fmfforce{1/2w,2/3h}{v7}
\fmfforce{1/2w,1/3h}{v8}
\fmf{double,width=0.2mm}{v1,v5}
\fmf{double,width=0.2mm}{v2,v6}
\fmf{double,width=0.2mm,right=0.3}{v3,v7}
\fmf{double,width=0.2mm,left=0.3}{v4,v8}
\fmf{double,width=0.2mm,left=1}{v5,v7,v5}
\fmfv{decor.size=0, label=${\scriptstyle 1}$, l.dist=1mm, l.angle=-135}{v1}
\fmfv{decor.size=0, label=${\scriptstyle 2}$, l.dist=1mm, l.angle=135}{v2}
\fmfv{decor.size=0, label=${\scriptstyle 5}$, l.dist=1mm, l.angle=0}{v3}
\fmfv{decor.size=0, label=${\scriptstyle 6}$, l.dist=1mm, l.angle=0}{v4}
\end{fmfgraph*} } } 
\hspace*{3mm}  
\dphidouble{
\parbox{14mm}{\centerline{
\begin{fmfgraph*}(9,3)
\setval
\fmfforce{0w,3/4h}{v1}
\fmfforce{3/9w,3/4h}{v2}
\fmfforce{6/9w,3/4h}{v3}
\fmfforce{1w,3/4h}{v4}
\fmfforce{1/2w,5/4h}{v5}
\fmfforce{1/2w,1/4h}{v6}
\fmf{double,width=0.2mm}{v1,v2}
\fmf{double,width=0.2mm}{v3,v4}
\fmf{double,width=0.2mm,left=1}{v5,v6,v5}
\fmfv{decor.size=0, label=${\scriptstyle 5}$, l.dist=0.5mm, l.angle=-180}{v1}
\fmfv{decor.size=0, label=${\scriptstyle 6}$, l.dist=0.5mm, l.angle=0}{v4}
\end{fmfgraph*} } }
}{3}{4} \hspace*{0.2cm} .
\\ && \nonumber 
\end{eqnarray}
Inserting (\ref{PELSFull1}) in (\ref{PELSFull2}) it turns out that the one-particle
irreducible four-point function also follows from\cite{PELSKonstantin}
\begin{eqnarray}
&& \nonumber \\
  \parbox{10mm}{\centerline{
  \begin{fmfgraph*}(6,6)
  \setval
  \fmfforce{0w,0h}{v1}
  \fmfforce{0w,1h}{v2}
  \fmfforce{1w,1h}{v3}
  \fmfforce{1w,0h}{v4}
  \fmfforce{1/3w,1/3h}{v5}
  \fmfforce{1/3w,2/3h}{v6}
  \fmfforce{2/3w,2/3h}{v7}
  \fmfforce{2/3w,1/3h}{v8}
  \fmf{double,width=0.2mm}{v1,v5}
\fmf{double,width=0.2mm}{v2,v6}
\fmf{double,width=0.2mm}{v3,v7}
\fmf{double,width=0.2mm}{v4,v8}
\fmf{double,width=0.2mm,left=1}{v5,v7,v5}
\fmfv{decor.size=0, label=${\scriptstyle 1}$, l.dist=1mm, l.angle=-135}{v1}
\fmfv{decor.size=0, label=${\scriptstyle 2}$, l.dist=1mm, l.angle=135}{v2}
\fmfv{decor.size=0, label=${\scriptstyle 3}$, l.dist=1mm, l.angle=45}{v3}
\fmfv{decor.size=0, label=${\scriptstyle 4}$, l.dist=1mm, l.angle=-45}{v4}
\end{fmfgraph*} } } 
%%%%
\hspace*{2mm} & = & \hspace*{3mm}
%%%%
  \parbox{7mm}{\centerline{
  \begin{fmfgraph*}(4.5,4.5)
  \setval
  \fmfforce{0w,0h}{v1}
  \fmfforce{0w,1h}{v2}
  \fmfforce{1w,1h}{v3}
  \fmfforce{1w,0h}{v4}
  \fmfforce{1/2w,1/2h}{v5}
  \fmf{double,width=0.2mm}{v1,v3}
  \fmf{double,width=0.2mm}{v2,v4}
\fmfv{decor.size=0, label=${\scriptstyle 1}$, l.dist=1mm, l.angle=-135}{v1}
\fmfv{decor.size=0, label=${\scriptstyle 2}$, l.dist=1mm, l.angle=135}{v2}
\fmfv{decor.size=0, label=${\scriptstyle 3}$, l.dist=1mm, l.angle=45}{v3}
\fmfv{decor.size=0, label=${\scriptstyle 4}$, l.dist=1mm, l.angle=-45}{v4}
\fmfdot{v5}
\end{fmfgraph*} } } 
%%%%
\hspace*{3mm} + \hspace*{2mm} \frac{1}{3} \hspace*{4mm}
%%%%
  \parbox{10mm}{\centerline{
  \begin{fmfgraph*}(8,8)
  \setval
  \fmfforce{0w,5.5/8h}{v1}
  \fmfforce{3/8w,5.5/8h}{v2}
  \fmfforce{1w,5.5/8h}{v3}
  \fmfforce{1w,1.5/8h}{v4}
  \fmfforce{1w,9.5/8h}{v5}
  \fmf{double,width=0.2mm}{v1,v2}
  \fmf{double,width=0.2mm}{v2,v3}
  \fmf{double,width=0.2mm,right=0.3}{v2,v4}
  \fmf{double,width=0.2mm,left=0.3}{v2,v5}
  \fmfv{decor.size=0, label=${\scriptstyle 1}$, l.dist=1mm, l.angle=-180}{v1}
  \fmfv{decor.size=0, label=${\scriptstyle 6}$, l.dist=1mm, l.angle=0}{v3}
  \fmfv{decor.size=0, label=${\scriptstyle 7}$, l.dist=1mm, l.angle=0}{v4}
  \fmfv{decor.size=0, label=${\scriptstyle 5}$, l.dist=1mm, l.angle=0}{v5}
  \fmfdot{v2}
  \end{fmfgraph*} } } 
%%%%
\hspace*{3mm} 
%%%%
\dphidouble{
  \parbox{11mm}{\centerline{
  \begin{fmfgraph*}(6,7)
  \setval
  \fmfforce{0w,2/7h}{v1}
  \fmfforce{0w,8/7h}{v2}
  \fmfforce{1w,8/7h}{v3}
  \fmfforce{1w,2/7h}{v4}
  \fmfforce{1/3w,4/7h}{v5}
  \fmfforce{1/3w,6/7h}{v6}
  \fmfforce{2/3w,6/7h}{v7}
  \fmfforce{2/3w,4/7h}{v8}
  \fmf{double,width=0.2mm}{v1,v5}
  \fmf{double,width=0.2mm}{v2,v6}
  \fmf{double,width=0.2mm}{v3,v7}
  \fmf{double,width=0.2mm}{v4,v8}
  \fmf{double,width=0.2mm,left=1}{v5,v7,v5}
  \fmfv{decor.size=0, label=${\scriptstyle 5}$, l.dist=0.5mm, l.angle=-180}{v1}
  \fmfv{decor.size=0, label=${\scriptstyle 2}$, l.dist=0.5mm, l.angle=180}{v2}
  \fmfv{decor.size=0, label=${\scriptstyle 3}$, l.dist=0.5mm, l.angle=0}{v3}
  \fmfv{decor.size=0, label=${\scriptstyle 4}$, l.dist=0.5mm, l.angle=0}{v4}
  \end{fmfgraph*} } } 
}{6}{7}
%%%%
\hspace*{3mm} + \hspace*{2mm} \frac{1}{6} \hspace*{2mm}
%%%%
  \parbox{10mm}{\centerline{
  \begin{fmfgraph*}(7,11)
  \setval
  \fmfforce{0w,13/11h}{v1}
  \fmfforce{2/7w,11/11h}{v2}
  \fmfforce{1w,13/11h}{v3}
  \fmfforce{2/7w,7/11h}{v4}
  \fmfforce{1/2w,5.5/11h}{v5}
  \fmfforce{3/7w,4.5/11h}{v6}
  \fmfforce{1/7w,6.5/11h}{v7}
  \fmfforce{1w,6.5/11h}{v8}
  \fmfforce{1w,3/11h}{v9}
  \fmf{double,width=0.2mm}{v1,v2}
  \fmf{double,width=0.2mm,left=0.1}{v2,v3}
  \fmf{double,width=0.2mm}{v2,v4}
  \fmf{double,width=0.2mm,right=0.5}{v2,v7}
  \fmf{double,width=0.2mm}{v5,v8}
  \fmf{double,width=0.2mm,right=0.1}{v6,v9}
  \fmf{double,width=0.2mm,left=1}{v7,v6,v7}
  \fmfv{decor.size=0, label=${\scriptstyle 1}$, l.dist=0.5mm, l.angle=135}{v1}
  \fmfv{decor.size=0, label=${\scriptstyle 5}$, l.dist=1mm, l.angle=0}{v3}
  \fmfv{decor.size=0, label=${\scriptstyle 6}$, l.dist=1mm, l.angle=25}{v8}
  \fmfv{decor.size=0, label=${\scriptstyle 7}$, l.dist=1mm, l.angle=0}{v9}
  \fmfdot{v2}
  \end{fmfgraph*} } } 
%%%%
\hspace*{3mm} 
%%%%
\dphidouble{
  \parbox{11mm}{\centerline{
  \begin{fmfgraph*}(6,7)
  \setval
  \fmfforce{0w,2/7h}{v1}
  \fmfforce{0w,8/7h}{v2}
  \fmfforce{1w,8/7h}{v3}
  \fmfforce{1w,2/7h}{v4}
  \fmfforce{1/3w,4/7h}{v5}
  \fmfforce{1/3w,6/7h}{v6}
  \fmfforce{2/3w,6/7h}{v7}
  \fmfforce{2/3w,4/7h}{v8}
  \fmf{double,width=0.2mm}{v1,v5}
  \fmf{double,width=0.2mm}{v2,v6}
  \fmf{double,width=0.2mm}{v3,v7}
  \fmf{double,width=0.2mm}{v4,v8}
  \fmf{double,width=0.2mm,left=1}{v5,v7,v5}
  \fmfv{decor.size=0, label=${\scriptstyle 5}$, l.dist=0.5mm, l.angle=-180}{v1}
  \fmfv{decor.size=0, label=${\scriptstyle 2}$, l.dist=0.5mm, l.angle=180}{v2}
  \fmfv{decor.size=0, label=${\scriptstyle 3}$, l.dist=0.5mm, l.angle=0}{v3}
  \fmfv{decor.size=0, label=${\scriptstyle 4}$, l.dist=0.5mm, l.angle=0}{v4}
  \end{fmfgraph*} } } 
}{6}{7}
%%%%
%\nonumber  \\ && \nonumber \\ 
%%%%                      %%%%%%%%%%%%%%%%%%%%
%&& 
\hspace*{0.2cm} + \hspace*{2mm} \frac{1}{2} \hspace*{3mm}
%%%%
  \parbox{14mm}{\centerline{
  \begin{fmfgraph*}(11,6)
  \setval
  \fmfforce{0w,1/6h}{v1}
  \fmfforce{0w,5/6h}{v2}
  \fmfforce{1w,1h}{v3}
  \fmfforce{1w,0h}{v4}
  \fmfforce{7/11w,1/3h}{v5}
  \fmfforce{7/11w,2/3h}{v6}
  \fmfforce{9/11w,2/3h}{v7}
  \fmfforce{9/11w,1/3h}{v8}
  \fmfforce{2/11w,1/2h}{v9}
  \fmf{double,width=0.2mm}{v1,v9}
  \fmf{double,width=0.2mm}{v2,v9}
  \fmf{double,width=0.2mm}{v3,v7}
  \fmf{double,width=0.2mm}{v4,v8}
  \fmf{double,width=0.2mm,left=1}{v5,v7,v5}
  \fmf{double,width=0.2mm,left=0.7}{v9,v6}
  \fmf{double,width=0.2mm,right=0.7}{v9,v5}
  \fmfv{decor.size=0, label=${\scriptstyle 1}$, l.dist=1mm, l.angle=-135}{v1}
  \fmfv{decor.size=0, label=${\scriptstyle 2}$, l.dist=1mm, l.angle=135}{v2}
  \fmfv{decor.size=0, label=${\scriptstyle 3}$, l.dist=1mm, l.angle=25}{v3}
  \fmfv{decor.size=0, label=${\scriptstyle 4}$, l.dist=1mm, l.angle=-25}{v4}
  \fmfdot{v9}
  \end{fmfgraph*} } } 
%%%%
+ \hspace*{2mm} \frac{1}{2} \hspace*{3mm}
%%%%                        
  \parbox{14mm}{\centerline{
  \begin{fmfgraph*}(11,6)
  \setval
  \fmfforce{0w,1/6h}{v1}
  \fmfforce{0w,5/6h}{v2}
  \fmfforce{1w,1h}{v3}
  \fmfforce{1w,0h}{v4}
  \fmfforce{7/11w,1/3h}{v5}
  \fmfforce{7/11w,2/3h}{v6}
  \fmfforce{9/11w,2/3h}{v7}
  \fmfforce{9/11w,1/3h}{v8}
  \fmfforce{2/11w,1/2h}{v9}
  \fmf{double,width=0.2mm}{v1,v9}
  \fmf{double,width=0.2mm}{v2,v9}
  \fmf{double,width=0.2mm}{v3,v7}
  \fmf{double,width=0.2mm}{v4,v8}
  \fmf{double,width=0.2mm,left=1}{v5,v7,v5}
  \fmf{double,width=0.2mm,left=0.7}{v9,v6}
  \fmf{double,width=0.2mm,right=0.7}{v9,v5}
  \fmfv{decor.size=0, label=${\scriptstyle 1}$, l.dist=1mm, l.angle=-135}{v1}
  \fmfv{decor.size=0, label=${\scriptstyle 3}$, l.dist=1mm, l.angle=135}{v2}
  \fmfv{decor.size=0, label=${\scriptstyle 4}$, l.dist=1mm, l.angle=25}{v3}
  \fmfv{decor.size=0, label=${\scriptstyle 2}$, l.dist=1mm, l.angle=-25}{v4}
  \fmfdot{v9}
  \end{fmfgraph*} } } 
%%%%
\nonumber  \\ && \nonumber \\ 
&&
+ \hspace*{2mm} \frac{1}{2} \hspace*{3mm}
%%%%
  \parbox{14mm}{\centerline{
  \begin{fmfgraph*}(11,6)
  \setval
  \fmfforce{0w,1/6h}{v1}
  \fmfforce{0w,5/6h}{v2}
  \fmfforce{1w,1h}{v3}
  \fmfforce{1w,0h}{v4}
  \fmfforce{7/11w,1/3h}{v5}
  \fmfforce{7/11w,2/3h}{v6}
  \fmfforce{9/11w,2/3h}{v7}
  \fmfforce{9/11w,1/3h}{v8}
  \fmfforce{2/11w,1/2h}{v9}
  \fmf{double,width=0.2mm}{v1,v9}
  \fmf{double,width=0.2mm}{v2,v9}
  \fmf{double,width=0.2mm}{v3,v7}
  \fmf{double,width=0.2mm}{v4,v8}
  \fmf{double,width=0.2mm,left=1}{v5,v7,v5}
  \fmf{double,width=0.2mm,left=0.7}{v9,v6}
  \fmf{double,width=0.2mm,right=0.7}{v9,v5}
  \fmfv{decor.size=0, label=${\scriptstyle 1}$, l.dist=1mm, l.angle=-135}{v1}
  \fmfv{decor.size=0, label=${\scriptstyle 4}$, l.dist=1mm, l.angle=135}{v2}
  \fmfv{decor.size=0, label=${\scriptstyle 2}$, l.dist=1mm, l.angle=25}{v3}
  \fmfv{decor.size=0, label=${\scriptstyle 3}$, l.dist=1mm, l.angle=-25}{v4}
  \fmfdot{v9}
  \end{fmfgraph*} } } 
%%%%
\hspace*{0.2cm} + \hspace*{2mm} \frac{1}{6} \hspace*{2mm}
%%%%
  \parbox{14mm}{\centerline{
  \begin{fmfgraph*}(11,11)
  \setval
  \fmfforce{1/11w,0h}{v1}
  \fmfforce{0w,1h}{v2}
  \fmfforce{3/11w,2/11h}{v3}
  \fmfforce{2/11w,9/11h}{v4}
  \fmfforce{3/11w,6.5/11h}{v5}
  \fmfforce{1.5/11w,7.5/11h}{v6}
  \fmfforce{4.5/11w,8.5/11h}{v7}
  \fmfforce{7/11w,4.5/11h}{v8}
  \fmfforce{7/11w,6.5/11h}{v9}
  \fmfforce{9.2/11w,6.7/11h}{v10}
  \fmfforce{9/11w,4.5/11h}{v11}
  \fmfforce{1w,8.5/11h}{v12}
  \fmfforce{1w,2.5/11h}{v13}
  \fmf{double,width=0.2mm}{v1,v3}
  \fmf{double,width=0.2mm}{v2,v4}
  \fmf{double,width=0.2mm}{v10,v12}
  \fmf{double,width=0.2mm}{v11,v13}
  \fmf{double,width=0.2mm,left=1}{v6,v7,v6}
  \fmf{double,width=0.2mm,left=1}{v9,v11,v9}
  \fmf{double,width=0.2mm}{v3,v5}
  \fmf{double,width=0.2mm,left=0.5}{v3,v6}
  \fmf{double,width=0.2mm,left=0.3}{v7,v9}
  \fmf{double,width=0.2mm,right=0.4}{v3,v8}
  \fmfv{decor.size=0, label=${\scriptstyle 1}$, l.dist=1mm, l.angle=-180}{v1}
  \fmfv{decor.size=0, label=${\scriptstyle 2}$, l.dist=1mm, l.angle=180}{v2}
  \fmfv{decor.size=0, label=${\scriptstyle 3}$, l.dist=1mm, l.angle=45}{v12}
  \fmfv{decor.size=0, label=${\scriptstyle 4}$, l.dist=1mm, l.angle=-45}{v13}
  \fmfdot{v3}
  \end{fmfgraph*} } } 
%%%%%
\hspace*{3mm} + \hspace*{2mm} \frac{1}{6} \hspace*{2mm}
%%%%
  \parbox{14mm}{\centerline{
  \begin{fmfgraph*}(11,11)
  \setval
  \fmfforce{1/11w,0h}{v1}
  \fmfforce{0w,1h}{v2}
  \fmfforce{3/11w,2/11h}{v3}
  \fmfforce{2/11w,9/11h}{v4}
  \fmfforce{3/11w,6.5/11h}{v5}
  \fmfforce{1.5/11w,7.5/11h}{v6}
  \fmfforce{4.5/11w,8.5/11h}{v7}
  \fmfforce{7/11w,4.5/11h}{v8}
  \fmfforce{7/11w,6.5/11h}{v9}
  \fmfforce{9.2/11w,6.7/11h}{v10}
  \fmfforce{9/11w,4.5/11h}{v11}
  \fmfforce{1w,8.5/11h}{v12}
  \fmfforce{1w,2.5/11h}{v13}
  \fmf{double,width=0.2mm}{v1,v3}
  \fmf{double,width=0.2mm}{v2,v4}
  \fmf{double,width=0.2mm}{v10,v12}
  \fmf{double,width=0.2mm}{v11,v13}
  \fmf{double,width=0.2mm,left=1}{v6,v7,v6}
  \fmf{double,width=0.2mm,left=1}{v9,v11,v9}
  \fmf{double,width=0.2mm}{v3,v5}
  \fmf{double,width=0.2mm,left=0.5}{v3,v6}
  \fmf{double,width=0.2mm,left=0.3}{v7,v9}
  \fmf{double,width=0.2mm,right=0.4}{v3,v8}
  \fmfv{decor.size=0, label=${\scriptstyle 1}$, l.dist=1mm, l.angle=-180}{v1}
  \fmfv{decor.size=0, label=${\scriptstyle 3}$, l.dist=1mm, l.angle=180}{v2}
  \fmfv{decor.size=0, label=${\scriptstyle 4}$, l.dist=1mm, l.angle=45}{v12}
  \fmfv{decor.size=0, label=${\scriptstyle 2}$, l.dist=1mm, l.angle=-45}{v13}
  \fmfdot{v3}
  \end{fmfgraph*} } } 
%%%%%%
\hspace*{3mm} + \hspace*{2mm} \frac{1}{6} \hspace*{2mm}
%%%%
  \parbox{14mm}{\centerline{
  \begin{fmfgraph*}(11,11)
  \setval
  \fmfforce{1/11w,0h}{v1}
  \fmfforce{0w,1h}{v2}
  \fmfforce{3/11w,2/11h}{v3}
  \fmfforce{2/11w,9/11h}{v4}
  \fmfforce{3/11w,6.5/11h}{v5}
  \fmfforce{1.5/11w,7.5/11h}{v6}
  \fmfforce{4.5/11w,8.5/11h}{v7}
  \fmfforce{7/11w,4.5/11h}{v8}
  \fmfforce{7/11w,6.5/11h}{v9}
  \fmfforce{9.2/11w,6.7/11h}{v10}
  \fmfforce{9/11w,4.5/11h}{v11}
  \fmfforce{1w,8.5/11h}{v12}
  \fmfforce{1w,2.5/11h}{v13}
  \fmf{double,width=0.2mm}{v1,v3}
  \fmf{double,width=0.2mm}{v2,v4}
  \fmf{double,width=0.2mm}{v10,v12}
  \fmf{double,width=0.2mm}{v11,v13}
  \fmf{double,width=0.2mm,left=1}{v6,v7,v6}
  \fmf{double,width=0.2mm,left=1}{v9,v11,v9}
  \fmf{double,width=0.2mm}{v3,v5}
  \fmf{double,width=0.2mm,left=0.5}{v3,v6}
  \fmf{double,width=0.2mm,left=0.3}{v7,v9}
  \fmf{double,width=0.2mm,right=0.4}{v3,v8}
  \fmfv{decor.size=0, label=${\scriptstyle 1}$, l.dist=1mm, l.angle=-180}{v1}
  \fmfv{decor.size=0, label=${\scriptstyle 4}$, l.dist=1mm, l.angle=180}{v2}
  \fmfv{decor.size=0, label=${\scriptstyle 2}$, l.dist=1mm, l.angle=45}{v12}
  \fmfv{decor.size=0, label=${\scriptstyle 3}$, l.dist=1mm, l.angle=-45}{v13}
  \fmfdot{v3}
  \end{fmfgraph*} } } \hspace*{0.2cm}. 
%%%%
\label{PELSFull3}
\end{eqnarray}
The graphical solution of the functional 
differential equations\index{functional differential equation} 
(\ref{PELSFull1}) and (\ref{PELSFull2}) or (\ref{PELSFull3}) leads to 
all one-particle irreducible diagrams\index{one-particle irreducible diagrams} 
which do not contain any line 
corrections. Once they are generated, we can recover the 
tadpole-free\index{tadpole} 
one-particle irreducible diagrams\index{one-particle irreducible diagrams} 
according to the following
algorithm\cite{PELSKonstantin}. 
At first we subtract the
one-loop correction  from the self-energy\index{self-energy} 
in order to obtain the 
tadpole-free\index{tadpole} self-energy\index{self-energy}
\begin{eqnarray}
\label{PELSRECOVER}
\parbox{11mm}{\centerline{
\begin{fmfgraph*}(9,3)
\setval
\fmfforce{0w,1/2h}{v1}
\fmfforce{3/9w,1/2h}{v2}
\fmfforce{6/9w,1/2h}{v3}
\fmfforce{1w,1/2h}{v4}
\fmfforce{1/2w,1h}{v5}
\fmfforce{1/2w,0h}{v6}
\fmf{wiggly,width=0.2mm}{v1,v2}
\fmf{wiggly,width=0.2mm}{v3,v4}
\fmf{dbl_wiggly,width=0.2mm,left=1}{v5,v6,v5}
\fmfv{decor.size=0, label=${\scriptstyle 1}$, l.dist=1mm, l.angle=-180}{v1}
\fmfv{decor.size=0, label=${\scriptstyle 2}$, l.dist=1mm, l.angle=0}{v4}
\end{fmfgraph*} } } 
\quad = \quad 
\parbox{11mm}{\centerline{
\begin{fmfgraph*}(9,3)
\setval
\fmfforce{0w,1/2h}{v1}
\fmfforce{3/9w,1/2h}{v2}
\fmfforce{6/9w,1/2h}{v3}
\fmfforce{1w,1/2h}{v4}
\fmfforce{1/2w,1h}{v5}
\fmfforce{1/2w,0h}{v6}
\fmf{double,width=0.2mm}{v1,v2}
\fmf{double,width=0.2mm}{v3,v4}
\fmf{double,width=0.2mm,left=1}{v5,v6,v5}
\fmfv{decor.size=0, label=${\scriptstyle 1}$, l.dist=1mm, l.angle=-180}{v1}
\fmfv{decor.size=0, label=${\scriptstyle 2}$, l.dist=1mm, l.angle=0}{v4}
\end{fmfgraph*} } } 
\hspace*{0.3cm} - \, \frac{1}{2} \hspace*{3mm}
\parbox{9mm}{\centerline{
\begin{fmfgraph*}(6,5)
\setval
\fmfforce{0w,0h}{v1}
\fmfforce{1/2w,0h}{v2}
\fmfforce{1w,0h}{v3}
\fmfforce{1/12w,1/2h}{v4}
\fmfforce{11/12w,1/2h}{v5}
\fmf{double,width=0.2mm}{v1,v3}
\fmf{double,width=0.2mm,left=1}{v4,v5,v4}
\fmfv{decor.size=0, label=${\scriptstyle 1}$, l.dist=1mm, l.angle=-180}{v1}
\fmfv{decor.size=0, label=${\scriptstyle 2}$, l.dist=1mm, l.angle=0}{v3}
\fmfdot{v2}
\end{fmfgraph*} } } \hspace*{0.2cm}.
\end{eqnarray}
Iterating the modified Dyson equation (\ref{PELSWIG2}), we then determine the
tadpole-free\index{tadpole} diagrams of the fully-interacting two-point 
function
$\mbox{\boldmath $G$}$ where the lines represent the
modified correlation function (\ref{PELSMDC}). 
Finally we insert this expansion into
the one-particle irreducible diagrams\index{one-particle irreducible diagrams}
of $\Sigma$ and $\Gamma$
determined from (\ref{PELSFull1}) and (\ref{PELSFull2}) or (\ref{PELSFull3}).
\section{Summary}
We have reviewed different approaches to graphically generate 
tadpole-free\index{tadpole}
one-particle irreducible diagrams\index{one-particle irreducible diagrams} 
together with their weights\index{weights} which
are needed for calculating the critical exponents\index{critical exponents} of 
$\phi^4$-theory\index{$\phi^4$-theory}. One
approach is based on the graphical
solution of Eqs. (\ref{PELSWIG1})--(\ref{PELSWIG3}), the latter being
more complex. The alternative approach 
consists of the simpler Eqs. (\ref{PELSFull1}) and (\ref{PELSFull2}) at the
expense of subsequent iterations of the modified Dyson equation 
(\ref{PELSWIG2}) with (\ref{PELSRECOVER}). In order to decide which of those
approaches is more efficient,
it is necessary to perform
further analytic studies or to automatize them by computer algebra.
\section*{Acknowledgments}
A.P. is deeply indebted to Professor Dr. Hagen 
Kleinert\index{KLEINERT, H.} for the
opportunity to work with him as a scientific assistant. His universal
physical knowledge, his many brilliant ideas and his truly encouraging
personality cause a thrilling scientific environment in his research
group. Both of us wish him a happy $60$th birthday. Finally we thank
Michael Bachmann, Dr.~Boris~Kastening, and Dr.~Bruno~van~den~Bossche
for sharing our interest in the recursive graphical construction of 
Feynman diagrams\index{Feynman diagrams}.
\end{fmffile}
\end{document}